\documentclass[lettersize,journal]{IEEEtran}
\usepackage{amsmath,amsfonts}      
\usepackage{bm}                    
\usepackage{textgreek}
\usepackage{algorithm,algpseudocode}
\algrenewcommand\Require{\textbf{Input: }}
\algrenewcommand\Ensure {\textbf{Output: }}

\usepackage[table,xcdraw]{xcolor}  
\usepackage{tikz}

\usepackage{array}
\usepackage{booktabs}              
\usepackage{multirow}
\usepackage{threeparttable}
\usepackage{tabularx}

\usepackage{colortbl}
\usepackage{pifont}
\usepackage{makecell}

\newcommand{\cmark}{\ding{51}}      
\newcommand{\xmark}{\ding{55}}      
\newcommand{\pmark}{\ensuremath{\triangle}} 

\newcommand{\mode}[1]{\texttt{#1}}

\usepackage{graphicx}
\usepackage[caption=false,font=footnotesize]{subfig}

\usepackage{textcomp}              

\usepackage{url}                   
\usepackage[hidelinks]{hyperref}
\usepackage{cite}                  

\usepackage{stfloats}              
\usepackage{cancel}

\begin{document}

\title{iDynamics: A Configurable Emulation Framework for Evaluating Microservice Scheduling Policies under Controllable Cloud--Edge Dynamics}

\author{Ming Chen, Muhammed Tawfiqul Islam, Maria Rodriguez Read, Rajkumar Buyya,~\IEEEmembership{Fellow,~IEEE}
    \thanks{Ming Chen, Muhammed Tawfiqul Islam, Maria Rodriguez Read, and Rajkumar Buyya are with the Quantum Cloud Computing and Distributed Systems (qCLOUDS) Laboratory, School of Computing and Information Systems, The University of Melbourne, Australia. (e-mail: mingc4@student.unimelb.edu.au; {tawfiqul.islam, maria.read, rbuyya}@unimelb.edu.au)}

}


\maketitle

\begin{abstract}

Designing and evaluating microservice scheduling policies in cloud--edge environments is challenging because application behavior is driven by multiple interacting dynamics: call-graph topology changes with the mix of request types, traffic along service chains becomes highly imbalanced, and cross-node network conditions such as latency and bandwidth fluctuate over time. Deploying and instrumenting large geo-distributed testbeds to study these effects is costly, disruptive, and often not repeatable. Existing simulators and emulators typically focus on either infrastructure dynamics or application workloads, and rarely capture their joint impact on end-to-end performance while running real services on a production-grade orchestration stack. This paper proposes \texttt{iDynamics}, a configurable emulation framework that exposes these dynamics as controllable experimental factors while running real microservice benchmarks on a Kubernetes-based cloud--edge cluster. \texttt{iDynamics} comprises three modular components, including Graph Dynamics Analyzer, Networking Dynamics Manager, and Scheduling Policy Extender. We use \texttt{iDynamics} to implement two policies as case studies demonstrating how the framework can be used to evaluate different policies under dynamic call graph evolutions and networking conditions. Experiments on a real cloud--edge cluster show that \texttt{iDynamics} accurately emulates targeted network conditions, tracks diverse call graph and traffic patterns, and helps quantify how different scheduling policies mitigate performance bottlenecks under controllable and repeatable dynamics.

\end{abstract}


\section{Introduction}

Microservice architectures have become the de facto style for building cloud and edge applications. By decomposing functionality into independently deployable services, they provide elasticity and rapid evolution of individual components. However, when tens or hundreds of microservices are deployed across a cloud--edge continuum, maintaining Service Level Agreements (SLAs) becomes non-trivial: user requests traverse long service chains, services are shared across applications, and traffic patterns vary significantly over time ~\cite{ms_survey1, Erms_ASPLOS23, SoCC2021_MS_luo}

In such settings, scheduling policies---that is, the placement and migration of microservice instances onto available nodes---must react to dynamics at three layers: (i) \textbf{workload dynamics}, such as changing mixes and proportions of request types; (ii) \textbf{application-level dynamics}, such as evolving call-graph structures and imbalanced traffic between pairs of services; (iii) \textbf{networking dynamics}, such as fluctuating cross-node delays and available bandwidth in the cloud--edge continuum. These intertwined effects directly impact the end-to-end performance of microservice applications deployed on Kubernetes~\cite{kubernetes} clusters.

Designing scheduling policies that cope with these dynamics already requires substantial effort; rigorously \emph{evaluating} them is even harder. Deploying alternative schedulers in production-scale, geographically distributed clusters is costly, disruptive, and rarely repeatable. Purely simulation-based evaluations, in contrast, often abstract away essential aspects of the Kubernetes stack, the service mesh, and real microservice implementations, which limits fidelity. Therefore, there is a need for an evaluation environment that runs real services while exposing key sources of non-determinism as experimental knobs rather than as uncontrolled environmental factors.


A large body of work has addressed microservice management and scheduling, for example through Service-Level Objective (SLO)-oriented resource management~\cite{FIRM_MS, Erms_ASPLOS23, grandslam_MS, Sage_ASPLOS21}, trace-based call graph analysis~\cite{TaskDependenceis_SoCC19, SoCC2021_MS_luo, Parslo_SoCC21}, network dynamics emulation~\cite{THUNDERSTORM_SRDS19, Kollaps_EuroSys20, SplayNet_TranNetworking16}, and network-aware pod placement in Kubernetes clusters~\cite{NetMARKS_InfoCOM21, ExtendingNetK8s_CCGRID22, ExtendingNetK8s_ICSOC22, OptTraffic_ICPP23}. Complementary work has proposed microservice simulators and performance-testing tools that couple real orchestrators with synthetic workloads. However, to the best of our knowledge, existing approaches either (i) focus on building new management algorithms and evaluate them in ad hoc environments, (ii) emulate network-level behavior without understanding microservice call graphs and SLAs, or (iii) provide simulators without a pluggable interface for alternative scheduling policies. None offers an integrated, open, Kubernetes-based framework that jointly (a) observes dynamic call graphs and traffic, (b) emulates heterogeneous cross-node latency and bandwidth, and (c) exposes a policy-agnostic interface for implementing and comparing schedulers under controllable dynamics.

To fill this gap, we propose \texttt{iDynamics}, a configurable emulation framework for evaluating microservice scheduling policies on Kubernetes-based cloud--edge clusters under controllable dynamics. In this work, we use the term \emph{controllable dynamics} to denote the capability to systematically emulate and vary call graph topologies, traffic loads along service chains, and cross-node network conditions, while monitoring the effects of these variations on end-to-end performance. Rather than proposing yet another scheduling algorithm, \texttt{iDynamics} focuses on the \emph{evaluation problem}: it instruments the service mesh, injects network dynamics, and provides an extensible policy interface so that both existing and new schedulers can be studied under repeatable conditions. The framework is implemented as a set of modular components, including the Graph Dynamics Analyzer, Networking Dynamics Manager, and Scheduling Policy Extender.

Our main contributions are as follows:
\begin{itemize}
    \item We propose the notion of \emph{controllable dynamics} for microservice scheduling evaluation and implement it across the call graph, traffic, and network layers in a cloud--edge cluster.
    \item We develop \texttt{iDynamics}, a modular emulation framework that combines the active call graph construction, destination-specific networking state emulation and validation, mixed workload and trace adapters for evaluation, and an executable scheduling-policy interface.
    \item We conduct extensive evaluations, including service mesh overhead, call graph construction overhead, validation of dynamic networking conditions, continuous call graph dynamics, executable example-policy studies with two representative microservice benchmarks, and a CPU-only MoE (Mixture of Experts)-style serving case study.
\end{itemize}

The rest of the paper is structured as follows. Section~\ref{sec:related} reviews related work. Section~\ref{sec:Motivation} discusses background and challenges. Section~\ref{sec:framework} introduces the \texttt{iDynamics} framework and its main components. Sections~\ref{sec: Graph_Dynamics_Analyzer}, \ref{sec: Networking_Dynamics_Manager}, and~\ref{sec: Policy_Extender} detail the designs of the Graph Dynamics Analyzer, Networking Dynamics Manager, and Scheduling Policy Extender, respectively. Section~\ref{sec:performance_evaluation} presents performance evaluation and case studies, and Section~\ref{sec:conclusion} concludes the paper and outlines future work.

\begin{table*}[!t]
\centering
\caption{Capability comparison of iDynamics with representative work.}
\label{tab:table_relatedWork}

\setlength{\tabcolsep}{2.2pt}
\renewcommand{\arraystretch}{1.12}

\resizebox{\textwidth}{!}{%
\begin{tabular}{l|ccc|ccc|ccc|cccc|ccc}
\hline\hline
\multirow{2}{*}{\textbf{Work}} &
\multicolumn{3}{c|}{\textbf{Microservice Management}} &
\multicolumn{3}{c|}{\textbf{Workload--Dynamics Control}} &
\multicolumn{3}{c|}{\textbf{Call--Graph Dynamics Analysis}} &
\multicolumn{4}{c|}{\textbf{Network--Dynamics Control}} &
\multicolumn{3}{c}{\textbf{Scheduling--Policy Extensibility}}
\\
\cline{2-17}
& \makecell{SLA/SLO\\Support}
& \makecell{Placement/\\Scheduling}
& \makecell{E2E\\Metrics}
& \makecell{Multiple\\Request Types}
& \makecell{Mixed\\Proportions}
& \makecell{Temporal\\Mix Modes}
& \makecell{Dependency\\Discovery}
& \makecell{Traffic\\Analysis}
& \makecell{Graph\\Evolution}
& Emulation
& Latency
& Bandwidth
& Measurement
& \makecell{Pluggable\\Interface}
& \makecell{Utility\\Modules}
& \makecell{Cloud--Edge\\Support}
\\
\hline\hline

FIRM \cite{FIRM_MS}
& \cmark & \cmark & \cmark
& \cmark & \pmark & \pmark
& \cmark & \pmark & \pmark
& \cmark & \cmark & \cmark & \xmark
& \xmark & \xmark & \xmark
\\ \hline

Erms \cite{Erms_ASPLOS23}
& \cmark & \cmark & \cmark
& \pmark & \xmark & \xmark
& \cmark & \xmark & \pmark
& \xmark & \xmark & \xmark & \xmark
& \xmark & \xmark & \xmark
\\ \hline

GrandSLAm \cite{grandslam_MS}
& \cmark & \cmark & \cmark
& \cmark & \xmark & \xmark
& \pmark & \xmark & \xmark
& \xmark & \xmark & \xmark & \xmark
& \xmark & \xmark & \xmark
\\ \hline

Sage \cite{Sage_ASPLOS21}
& \cmark & \cmark & \cmark
& \pmark & \xmark & \xmark
& \cmark & \pmark & \pmark
& \xmark & \xmark & \pmark & \cmark
& \xmark & \xmark & \xmark
\\ \hline

Kollaps \cite{Kollaps_EuroSys20}
& \xmark & \xmark & \pmark
& \xmark & \xmark & \xmark
& \xmark & \xmark & \xmark
& \cmark & \cmark & \cmark & \cmark
& \xmark & \xmark & \pmark
\\ \hline

NetMARKS \cite{NetMARKS_InfoCOM21}
& \pmark & \cmark & \pmark
& \xmark & \xmark & \xmark
& \cmark & \cmark & \pmark
& \xmark & \xmark & \xmark & \xmark
& \pmark & \xmark & \cmark
\\ \hline

OptTraffic \cite{OptTraffic_ICPP23}
& \xmark & \cmark & \pmark
& \xmark & \xmark & \xmark
& \pmark & \cmark & \pmark
& \xmark & \xmark & \xmark & \xmark
& \xmark & \xmark & \xmark
\\ \hline

THUNDERSTORM \cite{THUNDERSTORM_SRDS19}
& \xmark & \xmark & \pmark
& \xmark & \xmark & \xmark
& \xmark & \xmark & \xmark
& \cmark & \cmark & \cmark & \pmark
& \xmark & \xmark & \pmark
\\ \hline

MiSim \cite{MiSim_k8sLoop_simu_PerformanceTools2022}
& \pmark & \pmark & \cmark
& \cmark & \pmark & \cmark
& \xmark & \xmark & \xmark
& \pmark & \cmark & \xmark & \xmark
& \pmark & \pmark & \xmark
\\ \hline

CloudNativeSim \cite{CloudNativeSim_SPE25}
& \cmark & \cmark & \cmark
& \cmark & \cmark & \pmark
& \xmark & \xmark & \xmark
& \xmark & \xmark & \xmark & \xmark
& \cmark & \cmark & \xmark
\\ \hline

TraDE \cite{TraDE_TPDS2026}
& \cmark & \cmark & \cmark
& \cmark & \pmark & \xmark
& \cmark & \cmark & \pmark
& \cmark & \cmark & \xmark & \cmark
& \xmark & \xmark & \xmark
\\ \hline

\textbf{iDynamics (This work)}
& \cmark & \cmark & \cmark
& \cmark & \cmark & \cmark
& \cmark & \cmark & \cmark
& \cmark & \cmark & \cmark & \cmark
& \cmark & \cmark & \cmark
\\
\hline\hline
\end{tabular}
}

\par\vspace{1mm}
\parbox{\textwidth}{%
\footnotesize\raggedright
\emph{Legend:}
\cmark\ = full support; \pmark\ = partial or implicit support; \xmark\ = not reported.

}

\end{table*}

\section{Related Work}
\label{sec:related}

The dynamics of microservice-based systems in cluster and cloud--edge environments have been studied from multiple angles, including resource management, call graph analysis, network emulation, and network-aware scheduling. In addition, several frameworks and benchmarks support the simulation or emulation of microservice workloads. We briefly review these directions and position \texttt{iDynamics} relative to them. Table~\ref{tab:table_relatedWork} compares \texttt{iDynamics} with representative work using a capability-oriented coding rule: full support denotes a capability implemented or exposed by the proposed system itself, whereas partial support includes narrower, offline/manual, simulation-only, or evaluation-harness support.

\subsection{Microservice Management and Resource Scheduling}

FIRM~\cite{FIRM_MS} leverages online telemetry and machine learning models to localize SLO violations and resource contentions, then mitigates them through fine-grained reprovisioning actions. Erms~\cite{Erms_ASPLOS23} formulates scalable resource management models for shared microservices with large call graphs, assigning latency objectives to individual services. GrandSLAm~\cite{grandslam_MS} predicts completion times for stages of microservice execution and dynamically batches and reorders requests based on these predictions. Sage~\cite{Sage_ASPLOS21} uses probabilistic models to diagnose cascading QoS violations in interactive microservices at scale. TraDE~\cite{TraDE_TPDS2026} introduces a network- and traffic-aware adaptive scheduling framework to maintain QoS targets under changing workloads and network conditions. However,
these systems demonstrate sophisticated management logic but treat their execution environment largely as given: they do not expose a generic evaluation framework that allows different scheduling policies to be plugged in and compared under configurable call graph, traffic, and networking dynamics. 

\subsection{Network Dynamics Emulation}
Network dynamics strongly influence microservice performance in cloud--edge systems. THUNDERSTORM~\cite{THUNDERSTORM_SRDS19} evaluates distributed systems under dynamic topologies using micro- and macro-benchmarks. Kollaps~\cite{Kollaps_EuroSys20} provides a decentralized topology emulator, while SplayNet~\cite{SplayNet_TranNetworking16} builds on Splay~\cite{Splay} to emulate arbitrary overlay topologies. These tools are effective at reproducing network-level behavior but are agnostic primarily to microservice call graphs, SLAs, and scheduling policies.
At the level of individual nodes, several works inject delays or shape bandwidth using Linux traffic control primitives~\cite{FIRM_MS, delay_ms_TMC22, delay_edge_SPE22, Adaptive_ms_IPDPS21, adaptive_ms_cloudEdge_TPDS21}. Most, however, either employ static or uniform delay matrices or apply configurations at the granularity of a whole node, which makes it difficult to study heterogeneous pair-wise conditions in a cloud--edge cluster.  

\subsection{Network-Aware Microservice Scheduling}
Network-aware schedulers exploit infrastructure-level information when placing microservices. NetMARKS~\cite{NetMARKS_InfoCOM21} uses service-mesh metrics from Istio \cite{istio} to place Kubernetes pods for 5G edge applications. Marchese et al. extend the default Kubernetes scheduler with network-aware placement policies~\cite{ExtendingNetK8s_CCGRID22, ExtendingNetK8s_ICSOC22}. OptTraffic~\cite{OptTraffic_ICPP23} optimizes cross-machine traffic for multi-replica microservices by migrating containers with strong dependencies. However, these works confirm the value of incorporating network metrics into scheduling decisions but typically introduce a specific policy and evaluate it against the default Kubernetes scheduler, often in static or lightly controlled environments. 




\subsection{Microservice Simulators and Evaluation Tools}
Beyond schedulers and network emulators, several tools support simulation or emulation of microservice systems. Benchmark suites such as DeathStarBench~\cite{deathStarBench_ASPLOS19} provide realistic microservice applications and workloads. $\mu$Bench~\cite{muBench_TPDS23} benchmarks microservice applications with dummy services on Kubernetes-based platforms. MiSim~\cite{MiSim_k8sLoop_simu_PerformanceTools2022} models multiple service operations, time-varying workload generators, latency fault injection, resilience mechanisms, and extensible strategy implementations. More recently, CloudNativeSim~\cite{CloudNativeSim_SPE25} builds on CloudSim~\cite{CloudSim_SPE2011} to provide a discrete-event simulator for cloud-native, microservice-based applications, modeling service graphs, pods, and virtual machines and exposing policy interfaces for service management. However, these works provide essential building blocks and taxonomies but do not offer a unified, open framework that simultaneously (i) observes dynamic microservice call graphs, (ii) emulates heterogeneous cross-node network conditions, and (iii) exposes a pluggable interface for arbitrary scheduling policies while running real microservice containers on a Kubernetes-based cloud--edge cluster. \texttt{iDynamics} is designed to fill these gaps.

\begin{figure*}[!t]
\centering

\subfloat[
A workflow illustrating containerized microservice execution and communication under networking dynamics in a cloud--edge continuum.
\label{fig:cloud_edge_ms_dynamics}
]{
\includegraphics[width=0.31\textwidth]{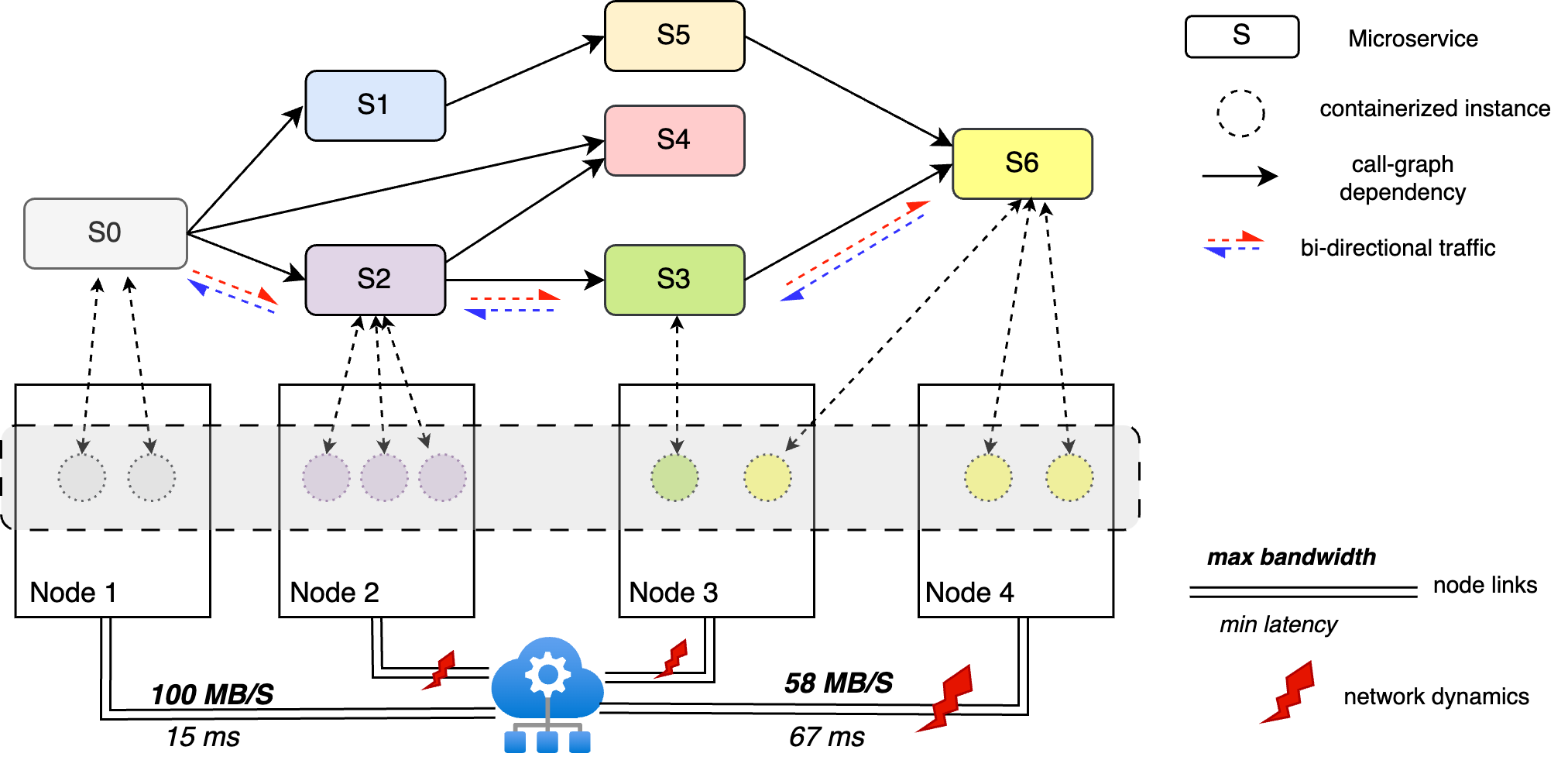}
}
\hfill
\subfloat[
Different request types trigger distinct call graph topologies. (application source \cite{deathStarBench_ASPLOS19})
\label{fig:dynamic_graph_socialNet}
]{
\includegraphics[width=0.31\textwidth]{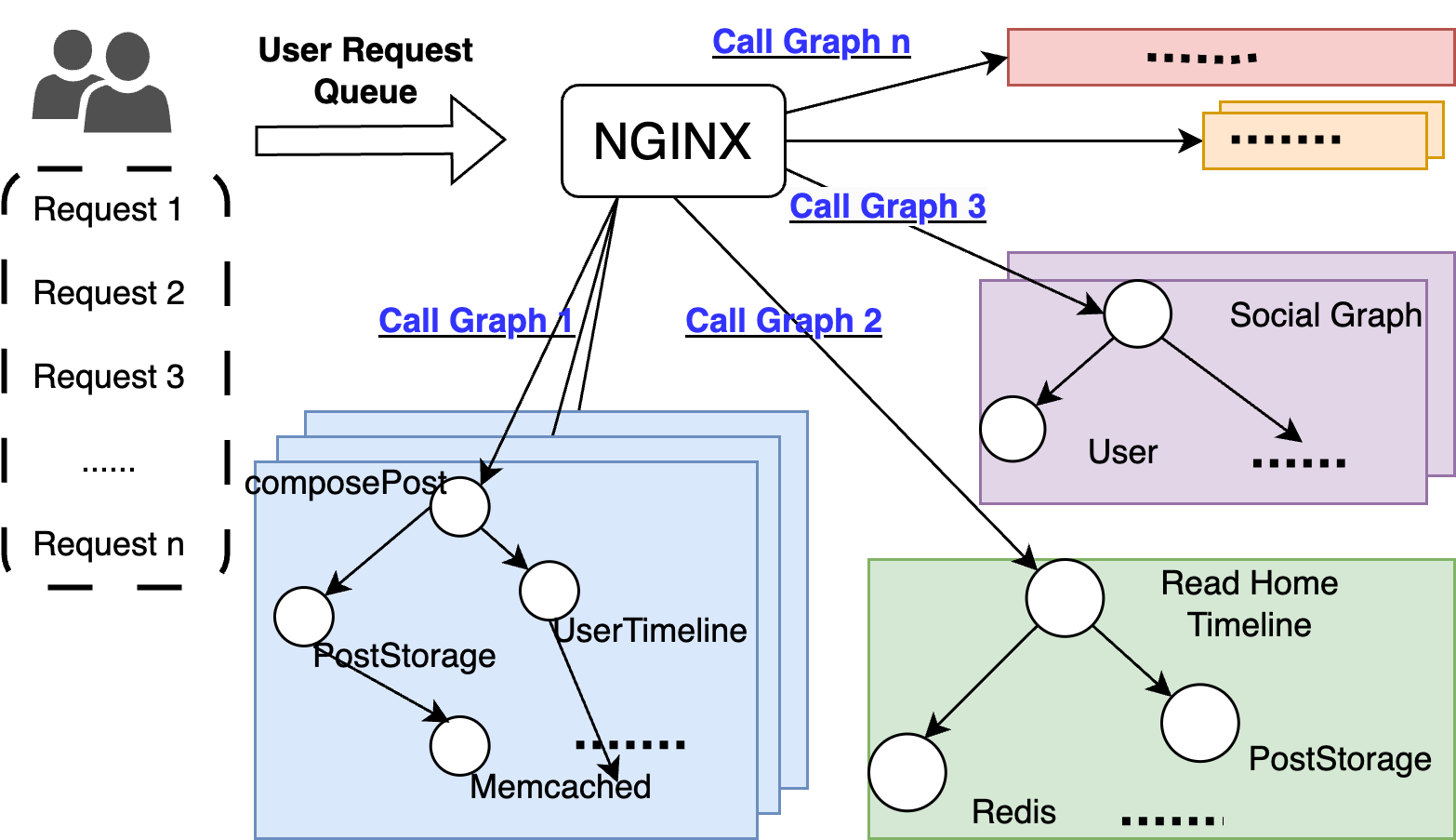}
}
\hfill
\subfloat[
A scenario of imbalanced traffic loads among UM--DM service pairs under high workload, 5k QPS. (application source \cite{deathStarBench_ASPLOS19})
\label{fig:motivation3_traffic}
]{
\includegraphics[width=0.31\textwidth]{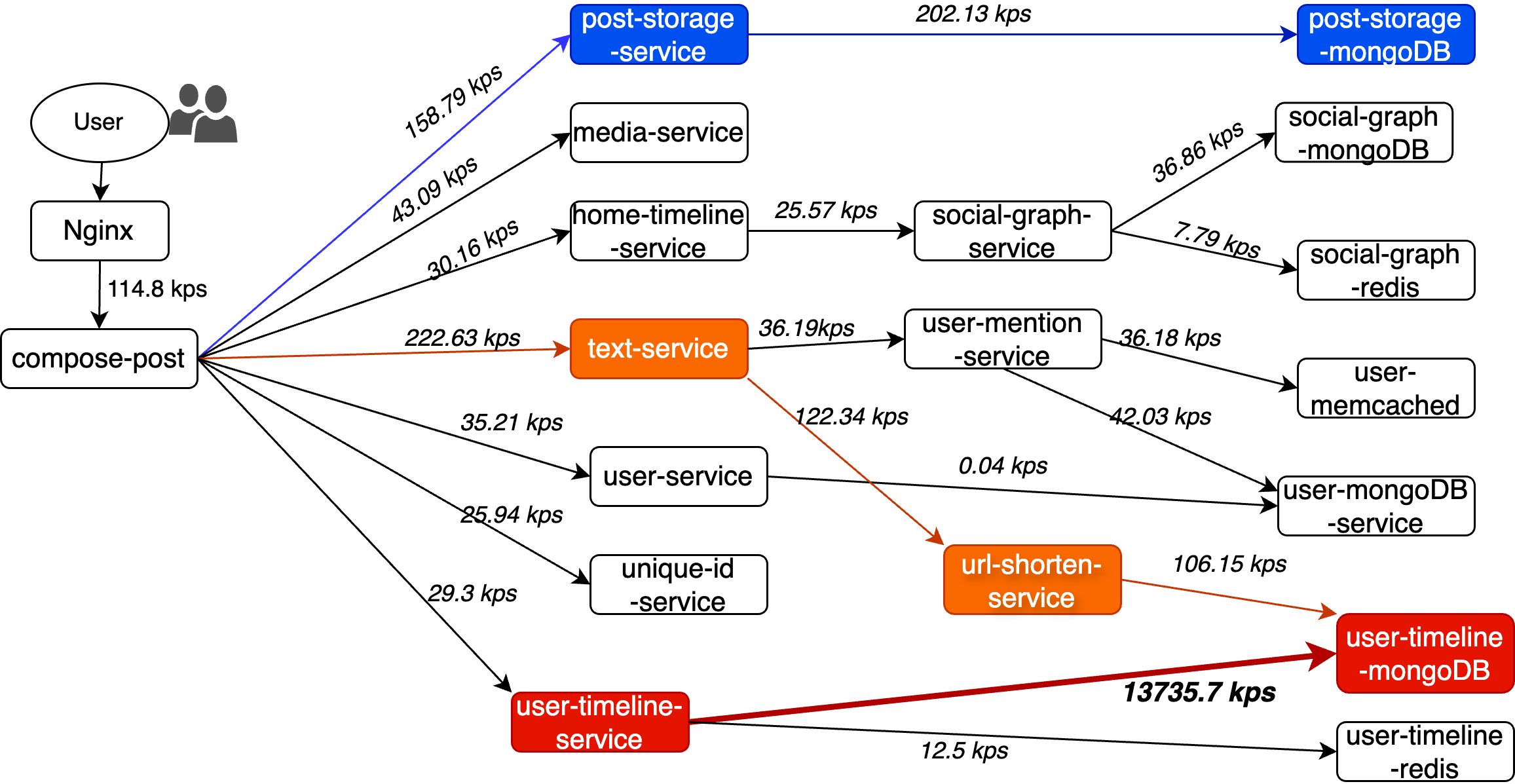}
}

\caption{Motivating examples for microservice scheduling in cloud--edge environments:
(a) network dynamics in the cloud--edge continuum,
(b) request-dependent call graph variability, and
(c) imbalanced traffic distribution across UM--DM pairs.}
\label{fig:motivation_overview}
\end{figure*}

\section{Background and Challenges}
\label{sec:Motivation}

\subsection{Background}
Microservice architecture decomposes an application into small, independently deployable services that communicate over the network \cite{SoCC2021_MS_luo, deathStarBench_ASPLOS19}. In cloud--edge computing, these services can be deployed in central cloud data centers, on edge nodes near end users and devices, or across both. In practice, an application may be partitioned so that some microservices run in the cloud (for global data, analytics, or coordination) while latency-sensitive components execute at the edge, communicating via Remote Procedure Calls (RPCs) or REST APIs.

This hybrid deployment improves responsiveness by reducing the distance that latency-critical data must travel and offloading work from the cloud to local nodes. At the same time, it increases sensitivity to network conditions and placement decisions: each network hop between microservices incurs latency and consumes bandwidth, and shared microservices often become hotspots as multiple application workflows converge on them. In the remainder of this paper, we use the term \emph{upstream microservice} (UM) to denote the caller in a dependency pair and \emph{downstream microservice} (DM) to denote the callee. Scheduling decisions determine where microservice pods are placed on cluster nodes and under what conditions pods are migrated between nodes.

\subsection{Challenges}


Figure~\ref{fig:cloud_edge_ms_dynamics} illustrates an envisioned workflow of microservices executing across multiple nodes under dynamic network conditions in a cloud--edge continuum. Building effective scheduling policies and rigorously evaluating them requires addressing at least three major challenges.

\subsubsection{Dynamic Networking Conditions}
In cloud--edge scenarios, microservice applications often experience dynamic, heterogeneous networking conditions. Cross-node delays and available bandwidth can vary due to contention, background traffic, and routing changes. As a result, fluctuating cross-node latencies and bandwidth constraints significantly affect end-to-end response times and throughput. Evaluating how scheduling policies behave under such conditions requires a framework that can emulate and measure realistic, time-varying network profiles, rather than relying solely on the low-latency conditions of a laboratory cluster.

\subsubsection{Variability in Call Graph Topologies}


Microservice requests traverse complex call graphs, which vary depending on the types of incoming requests. As shown in Figure~\ref{fig:dynamic_graph_socialNet}, widely used benchmark applications such as Social Network \cite{deathStarBench_ASPLOS19} expose different request types (e.g., compose-post, read-home-timeline, read-user-timeline), each triggering a distinct call graph structure. The relative proportions of these request types change over time, leading to evolving mixtures of call graphs and shifting bottlenecked edges. Scheduling policies must therefore handle both topology variability (which services are involved) and workload variability (how frequently each path is exercised). 

\subsubsection{Imbalanced Traffic Distribution}
Even within a fixed call graph topology, traffic loads across UM--DM pairs can be highly imbalanced. Some service pairs carry orders of magnitude more requests or larger payloads than others, and this imbalance becomes more pronounced under high load. Figure~\ref{fig:motivation3_traffic} shows an example from the Social Network application where a small subset of UM--DM pairs accounts for the majority of traffic at 5k queries per second (QPS). The pair \texttt{user-timeline-service} $\rightarrow$ \texttt{user-timeline-mongoDB}, for instance, experiences much higher traffic than other edges in the same call graph.

Such imbalances can cause localized bottlenecks that dominate end-to-end latency, especially when the corresponding microservice pods are placed on nodes connected by high-latency or low-bandwidth links. Evaluating scheduling policies in this context requires a framework that (i) observes bi-directional traffic between UM--DM pairs, (ii) identifies imbalanced edges, and (iii) can manipulate network conditions to stress specific parts of the call graph.

These challenges motivate the need for an evaluation framework that can jointly observe and control workload, application, and networking dynamics when studying microservice scheduling policies.


\section{iDynamics Framework}

\label{sec:framework}
\begin{figure*}[t]
    \centering
    \includegraphics[width=0.85\textwidth]{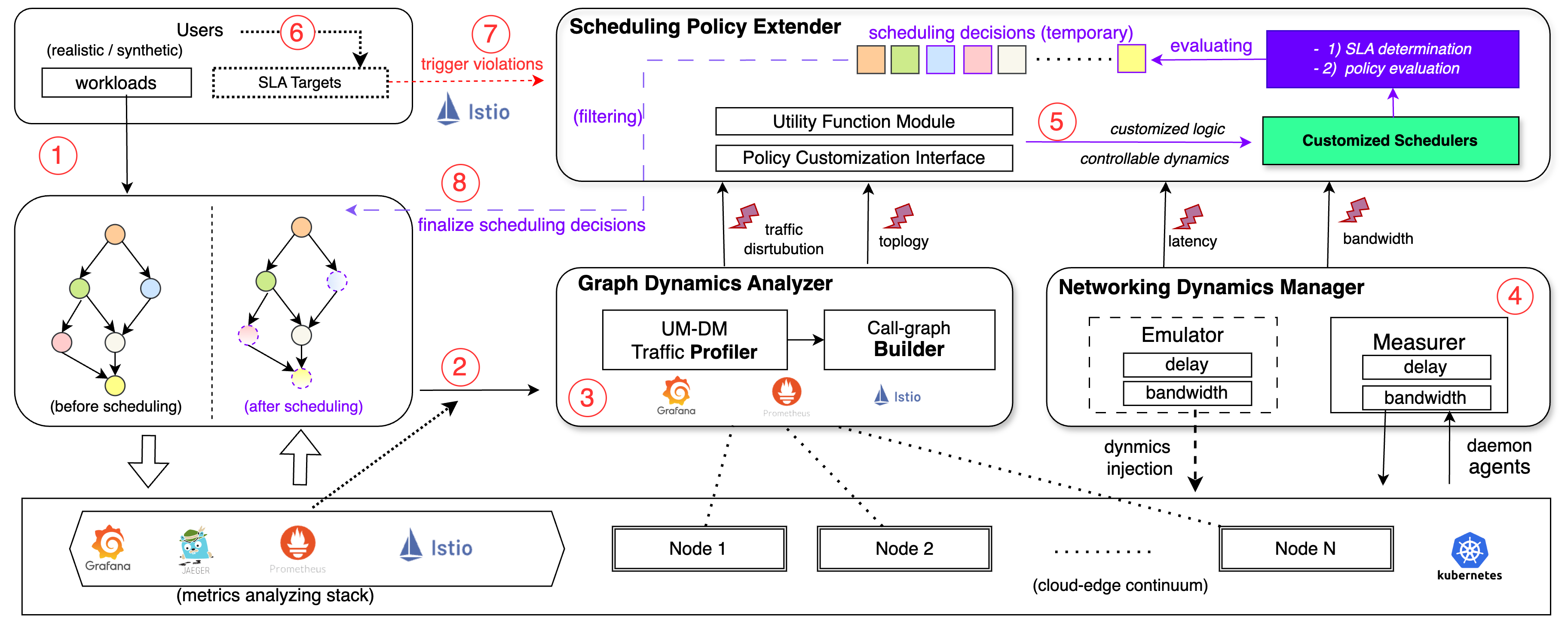}
    \caption{\texttt{iDynamics} framework architecture and working procedure.}
    \label{fig:iDynamics_framework}
\end{figure*}

Motivated by the challenges discussed above, we propose \texttt{iDynamics}, a unified emulation framework for implementing and evaluating microservice scheduling policies in cloud--edge Kubernetes clusters. Specifically, \texttt{iDynamics} is designed to: (i) run real containerized microservices and workloads; (ii) expose call graph, traffic, and network conditions as controllable experimental factors; and
(iii) provide a pluggable interface where different scheduling policies can be implemented, executed, and compared. Figure~\ref{fig:iDynamics_framework} presents the architecture of \texttt{iDynamics} and the interactions among its components during evaluation and scheduling. The framework supports managing diverse dynamics and implementing scheduling policies across the cloud--edge continuum. It currently targets Kubernetes and Istio, but the design principles generalize to other orchestrators and service meshes.

The key components of \texttt{iDynamics} are:
\begin{itemize}
    \item \textbf{Graph Dynamics Analyzer}: This component consists of a UM--DM Traffic Profiler and a Call-Graph Builder. It reconstructs the call graph topology triggered by different request types and quantifies bi-directional traffic between microservices (and their replicas) under varying workloads (Section~\ref{sec: Graph_Dynamics_Analyzer}).
    \item \textbf{Networking Dynamics Manager}: This component includes an Emulator and a distributed Measurer. The Emulator generates configurable cross-node latency and bandwidth patterns using Linux traffic control, while the Measurer collects actual communication metrics between nodes via lightweight daemon agents deployed across the cloud--edge cluster (Section~\ref{sec: Networking_Dynamics_Manager}).
    \item \textbf{Scheduling Policy Extender}: This component provides a Policy Customization Interface and a Utility Function Module. It enables rapid development and evaluation of scheduling strategies by exposing a policy-agnostic abstraction of nodes, pods, and metrics, and by providing helper functions to access cluster and network state (Section~\ref{sec: Policy_Extender}).
\end{itemize}

As depicted in Figure~\ref{fig:iDynamics_framework}, the working procedure of \texttt{iDynamics} consists of multiple stages:
\begin{enumerate}
    \item[\textcircled{\raisebox{-0.9pt}{1}}] Users submit realistic or synthetic workloads (with different request types and queries per second) to the deployed microservice applications.
    \item[\textcircled{\raisebox{-0.9pt}{2}}-\textcircled{\raisebox{-0.9pt}{3}}] Performance metrics are collected using a monitoring stack (e.g., Istio \cite{istio}, Jaeger\cite{Jaeger_tracing}, Prometheus \cite{Prometheus}). The Graph Dynamics Analyzer then builds the triggered call graph and analyzes traffic between microservices, forwarding these insights to the Scheduling Policy Extender.
    \item[\textcircled{\raisebox{-0.9pt}{4}}] In parallel, the Networking Dynamics Manager measures real-time cross-node conditions and can optionally inject dynamic delay and bandwidth patterns into the cloud--edge environment.
    \item[\textcircled{\raisebox{-0.9pt}{5}}] Using the collected dynamics (call graphs, traffic distributions, and node-level latency and bandwidth), users implement and evaluate customized scheduling policies via the policy interface and utility module. In this context, a \emph{scheduling decision} is a mapping of one or more pods to target nodes, possibly involving migrations to new nodes.
    \item[\textcircled{\raisebox{-0.9pt}{6}}] Users specify SLA targets (e.g., tail latency thresholds) that determine when a policy should be triggered.
    \item[\textcircled{\raisebox{-0.9pt}{7}}] SLA violations for policy evaluation are induced either by tightening SLA criteria or by injecting more intense dynamics (e.g., increased delay or workload), causing performance degradation. Candidate scheduling decisions produced by the policy are stored in a decision queue and can be filtered based on compliance requirements or constraints on which services are allowed to migrate.
    \item[\textcircled{\raisebox{-0.9pt}{8}}] Finally, the filtered scheduling decisions are executed by reconfiguring pod placements in the Kubernetes cluster, adjusting the deployment according to current conditions.
\end{enumerate}
With \texttt{iDynamics}, scheduling policies can thus be implemented, tested, and refined under realistic yet controllable conditions. For example, traffic- and latency-aware scheduling strategies can be developed using dynamic metrics from the Graph Dynamics Analyzer, emulated network dynamics from the Networking Dynamics Manager, and the extensible interfaces provided by the Scheduling Policy Extender.

\section{Graph Dynamics Analyzer}
\label{sec: Graph_Dynamics_Analyzer}

Considering the characteristics of microservice call graphs described in
Section~\ref{sec:Motivation}, the \textit{Graph Dynamics Analyzer} is
designed to analyze (i) topology dynamics, i.e., variations in triggered
call graphs, and (ii) traffic dynamics, i.e., imbalanced traffic
distribution, of microservice applications deployed in the cloud--edge
continuum. To capture these dynamics, \texttt{iDynamics} leverages a
service mesh and implements two core modules: the \textit{UM--DM Traffic
Profiler} and the \textit{Call-graph Builder}, as illustrated in
Figure~\ref{fig:iDynamics_framework}.

\subsection{Service Mesh}

A service mesh enables near-real-time monitoring of service-to-service
traffic across microservice instances. A key requirement of the
\textit{Graph Dynamics Analyzer} is to capture and aggregate
bidirectional traffic between dependent microservices efficiently. When
there is only a single instance of each microservice, low-level traffic
statistics can be collected from the node operating system. However,
modern microservice applications typically scale individual services
horizontally. Multiple upstream microservice (UM) and downstream
microservice (DM) replicas introduce many-to-many communication patterns,
making naive per-instance traffic analysis expensive and fragile. To
address this, \texttt{iDynamics} adopts a service-mesh-based approach
that offloads traffic collection and aggregation to sidecar proxies.

Although the current system uses Istio, the Graph
Dynamics Analyzer only requires telemetry records that expose source
workload, destination workload, and traffic-volume counters. Other
service meshes or observability stacks can therefore be supported by exporting the same UM--DM traffic schema.

\subsubsection{Istio Service-Mesh Implementation}
To obtain fine-grained bidirectional traffic metrics between UM and DM
microservices and their replicas, we implement the \textit{UM--DM Traffic
Profiler} using the Istio service mesh~\cite{istio}. Istio is deployed as
an infrastructure layer that transparently intercepts service-to-service
traffic via sidecar proxies. In a Kubernetes cluster with Istio enabled,
each microservice pod contains the application container and an Envoy
sidecar container.  

\subsubsection{Overhead Analysis}
Introducing a service mesh adds an additional proxy layer
on the request path. We therefore measured its cost directly using a
controlled two-service Fortio benchmark~\cite{fortio_benchmark}.
Table~\ref{tab:istio_mesh_overhead} reports the measured overhead of
Istio sidecar injection. Across the three cluster scales, the no-sidecar
configuration achieved p95 latencies of 1.825--1.880~ms, while the
sidecar-enabled configuration increased p95 latency to 6.123--7.021~ms.
This corresponds to an additional latency of 4.299--5.141~ms per request.
Despite the additional proxy hop, throughput remained effectively
unchanged at approximately 100~rps, with throughput reductions below
0.04~rps in all cases. Prometheus/cAdvisor monitoring further showed that
the combined client/server sidecars consumed 0.033--0.043 CPU cores and
66.9--70.0~MiB of memory. These results indicate that
Istio introduces a small but measurable latency penalty and modest
resource overhead in our experimental environment.

\begin{table}[t]
\centering
\caption{Istio sidecar overhead in a controlled Fortio benchmark.}
\label{tab:istio_mesh_overhead}
\scriptsize
\setlength{\tabcolsep}{2pt}
\begin{threeparttable}
{
\begin{tabular}{@{}c l r r r r r r r@{}}
\toprule
\shortstack{Cluster\\scale} &
\shortstack{Config.} &
\shortstack{Thr.\\(rps)} &
\shortstack{$\Delta$Thr.\\(rps)} &
\shortstack{p95\\(ms)} &
\shortstack{$\Delta$p95\\(ms)} &
\shortstack{p99\\(ms)} &
\shortstack{CPU\\(cores)} &
\shortstack{Mem.\\(MiB)} \\
\midrule
5  & No-SC    & 99.988 & --     & 1.825 & --     & 2.308 & 0.000 & 0.0 \\
5  & Istio-SC & 99.961 & -0.027 & 6.123 & +4.299 & 6.971 & 0.033 & 66.9 \\
\addlinespace[1pt]
20 & No-SC    & 99.985 & --     & 1.876 & --     & 2.188 & 0.000 & 0.0 \\
20 & Istio-SC & 99.953 & -0.032 & 6.944 & +5.069 & 8.631 & 0.039 & 70.0 \\
\addlinespace[1pt]
45 & No-SC    & 99.987 & --     & 1.880 & --     & 2.411 & 0.000 & 0.0 \\
45 & Istio-SC & 99.955 & -0.032 & 7.021 & +5.141 & 7.613 & 0.043 & 68.8 \\
\bottomrule
\end{tabular}
}
\begin{tablenotes}[flushleft]
\footnotesize
\item[] \emph{Notes:} No-SC and Istio-SC denote no-sidecar
and Istio-sidecar configurations, respectively. Values are means over five
valid repetitions using Fortio~1.69.3 with 100~qps, 16 connections, 15~s
duration, and 1024-byte payloads. Deltas are relative to the corresponding
No-SC condition at the same worker-only placement-pool size. CPU and memory
report the additional combined client/server Istio sidecar footprint
measured using Prometheus.
\end{tablenotes}
\end{threeparttable}
\end{table}

\subsection{Call-graph Builder}
\label{subsec:call_graph_builder}

The \textit{Graph Dynamics Analyzer} periodically constructs an \emph{active runtime call graph} for each application namespace. The graph is active because it is derived from traffic observed during a monitoring interval, rather than from static source-code
dependencies. This distinction is important for microservice systems: production trace studies show that microservice call graphs are heavy-tailed, often tree-like, contain hotspot services, and can change topology across different request types or runtime contexts ~\cite{SoCC2021_MS_luo}. Accordingly, \texttt{iDynamics} treats call-graph topology and traffic stress as runtime experimental factors for evaluating scheduling policies under SLA targets.

Let \(I_t=(t-\Delta t,t]\) denote a monitoring interval and let \(\mathcal{M}_t\) be the set of running microservices in the namespace during that interval, with \(M_t=|\mathcal{M}_t|\). For an ordered upstream--downstream service pair \((\mu,\sigma)\), where
\(\mu,\sigma\in\mathcal{M}_t\) and \(\mu\neq\sigma\), the service mesh provides traffic counters grouped by source and destination labels. In this component implementation, \(\mu\) denotes the upstream microservice (UM), \(\sigma\) denotes the downstream microservice (DM), and the edge
direction is determined by the source--destination label pair. The sent and received byte counters are used to quantify edge weight; they do not create a reverse edge unless the reverse source--destination pair is also observed.

\textbf{Stress Element.}
A Stress Element (SE) models the observed runtime interaction between an upstream microservice and a downstream microservice. Let \(B^{\mathrm{sent}}_{\mu,\sigma}(I_t)\) and
\(B^{\mathrm{recv}}_{\mu,\sigma}(I_t)\) denote the bytes sent and received, respectively, for the labelled pair \((\mu,\sigma)\) during \(I_t\). We define the traffic stress of this pair as the mean bidirectional byte rate:
\begin{equation}
\label{eq:stress_SE}
\operatorname{stress}_{t}(\mu^{UM},\sigma^{DM}) =
\frac{
B^{\mathrm{sent}}_{\mu,\sigma}(I_t)
+
B^{\mathrm{recv}}_{\mu,\sigma}(I_t)
}{2\Delta t}.
\end{equation}

In \texttt{iDynamics}, these quantities are computed from Prometheus metrics such as istio\_tcp\_sent\_bytes\_total and istio\_tcp\_received\_bytes\_total, aggregated over
source workload and destination workload labels. A Stress Element is represented as
\[
SE_t(\mu^{UM},\sigma^{DM}) =
(\mu,\sigma,\operatorname{stress}_{t}(\mu,\sigma)).
\]

The active edge set during \(I_t\) is therefore
\begin{equation}
\label{eq:active_edge_set}
E_t =
\left\{
(\mu,\sigma)
\;\middle|\;
\mu,\sigma\in\mathcal{M}_t,\;
\mu\neq\sigma,\;
\operatorname{stress}_{t}(\mu,\sigma)>0
\right\}.
\end{equation}

The Call-graph Builder returns a directed weighted graph
\[
G_t=(\mathcal{M}_t,E_t,w_t),
\]
where
\(w_t(\mu,\sigma)=\operatorname{stress}_{t}(\mu,\sigma)\). Thus, vertices correspond to microservices and edges correspond to observed non-zero UM--DM communication during the interval. Absence of an edge means that no service-mesh traffic was observed for the pair during
\(I_t\); it does not rule out an inactive static code-level dependency.


\begin{algorithm}[t]
\caption{Build call graph with dependency topology and associated traffic stress.}
\label{alg:call_graph_builder}

\begin{algorithmic}[1]
\Statex \textbf{Input:} Application namespace \(\mathit{ns}\), time interval \(\Delta t\)
\Statex \textbf{Output:} Directed sparse call graph \(G\) with weighted traffic edges
\State Initialize \(G\) as an empty directed graph
\State \(\mathit{MS} \gets \textsc{GetRunningMS}(\mathit{ns})\)
\Comment{all running microservices in \(\mathit{ns}\)}
\State \((R,S) \gets \textsc{QueryMeshTrafficVectors}(\mathit{ns},\Delta t)\)
\Comment{received/sent bytes grouped by source and destination labels}

\ForAll{observed label pairs \((src,dst) \in \operatorname{keys}(R)\cup\operatorname{keys}(S)\)}
    \If{\((src \in \mathit{MS})\) and \((dst \in \mathit{MS})\) and \((src \neq dst)\)}
        \State \(r \gets \textsc{GetOrZero}(R,src,dst)\)
        \State \(s \gets \textsc{GetOrZero}(S,src,dst)\)
        \State \(\mathit{traffic} \gets (r+s)/(2\Delta t)\)
        \If{\(\mathit{traffic} > 0\)}
            \State \(\textsc{AddEdge}(G,src,dst,\mathit{traffic})\)
        \EndIf
    \EndIf
\EndFor

\State \Return \(G\)
\end{algorithmic}
\end{algorithm}

A dense all-pairs telemetry call graph builder enumerates the full
\(M_t(M_t-1)\) adjacency space even though most candidate pairs may have zero traffic. By contrast, Algorithm~\ref{alg:call_graph_builder} constructs the active graph in a sparse way, which directly aggregates service-mesh traffic vectors. It first retrieves sent and received traffic vectors for the namespace and interval, and then materializes only observed non-zero source--destination label pairs. This design is exact with respect to the monitoring interval: every positive traffic pair returned by the aggregate telemetry vectors is materialized as an edge, while zero-traffic pairs are not explicitly stored. For \(M_t\) services and \(|E_t|\) active edges, the
\textit{dense} call graph construction method performs \(O(M_t^2)\) pair checks and, in our measured implementation, issues \(2M_t(M_t-1)\) pair-wise Prometheus queries. In contrast, the \textit{sparse} method uses two aggregate Prometheus queries and \(O(|E_t|)\) edge materialization after the aggregate vectors are returned. If a workload genuinely activates most service pairs, then \(|E_t|\) approaches \(M_t(M_t-1)\) and materialization becomes quadratic; however, the telemetry query load remains constant in the number of aggregate Prometheus queries.

\begin{figure}[t]
    \centering
    \includegraphics[width=0.85\linewidth]{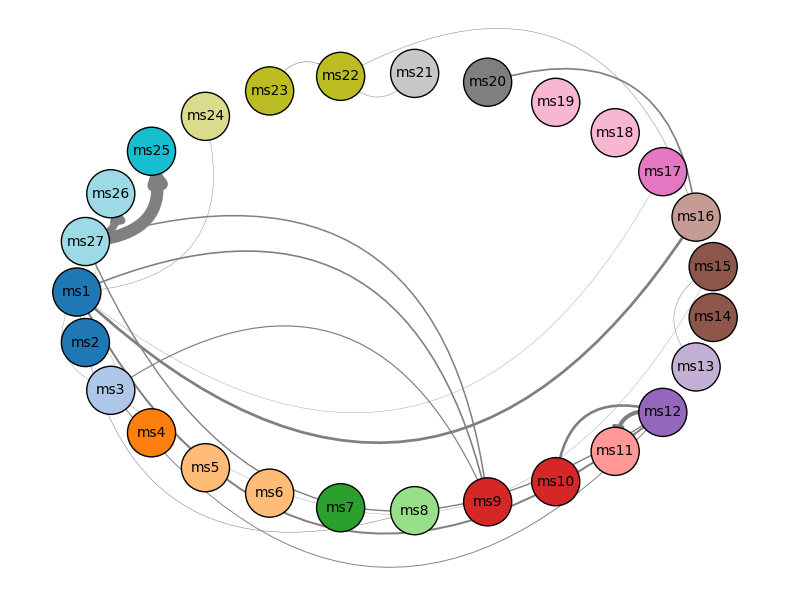}
    \caption{Based on Algorithm~\ref{alg:call_graph_builder}, the Graph Dynamics Analyzer builds a snapshot of the active call graph for an application with 27 microservices. The call graph shows dependencies and traffic flows.}
    \label{fig:graph_builder_demo}
\end{figure}

For each monitoring interval, the constructed call graph captures both topological dynamics and traffic dynamics. In particular, traffic-aware policies can co-locate or place low-latency paths between high-stress UM--DM pairs, while hybrid policies can combine call-graph stress with the network delay and bandwidth state exposed by the Networking Dynamics Manager. Figure~\ref{fig:graph_builder_demo} illustrates a call graph snapshot for a 27-service Social Network~\cite{deathStarBench_ASPLOS19} and is used as an explanatory visualization rather than just as numerical measurements. \textbf{As \(G_t\) is reconstructed periodically, overlapping request mixes are represented naturally}: the active graph for an interval contains the union of the UM--DM pairs exercised during that interval, and their weights reflect the aggregate traffic stress contributed by those requests. Thus, \texttt{iDynamics} is not limited to hard switching between pre-defined static call graphs; it can also expose gradual changes in edge weights and active dependencies as the request mix evolves.

\section{Networking Dynamics Manager}
\label{sec: Networking_Dynamics_Manager}

The \textit{Networking Dynamics Manager} provides two core capabilities in \texttt{iDynamics}: (i) emulation of controllable cross-node networking conditions and (ii) accurate measurement of those conditions in the cloud--edge continuum. It consists of two main components: the \textbf{Emulator} and the \textbf{Measurer}. Together, they enable systematic evaluation of scheduling policies under diverse delay and bandwidth configurations without requiring intrusive changes to production environments.


\begin{figure*}[t]
  \centering
  \subfloat[Packet egress with classful disciplines.\label{fig:enqueueDequeue}]{
    \includegraphics[width=.32\linewidth]{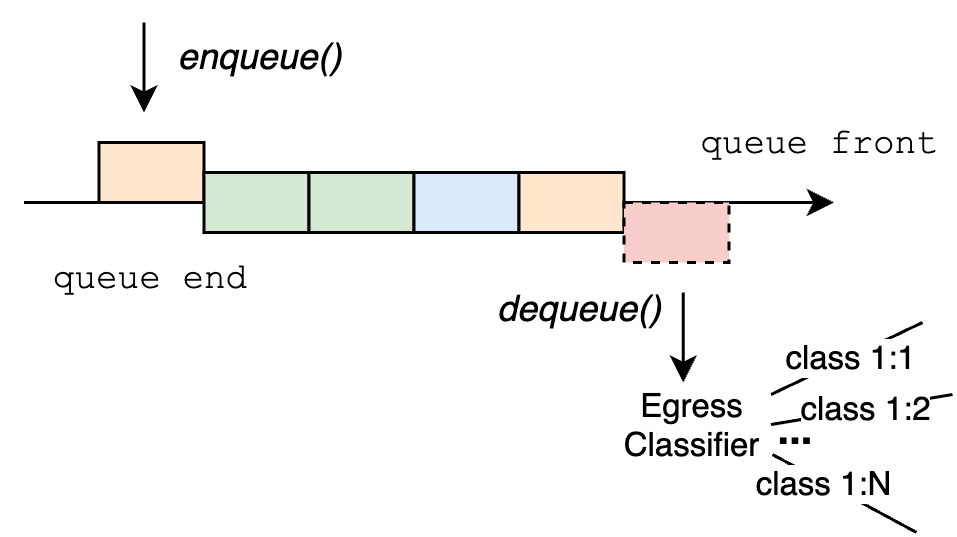}}
  \hfill
  \subfloat[Classful \texttt{qdisc} for delay emulation.\label{fig:qdisc_delay}]{
    \includegraphics[width=.32\linewidth]{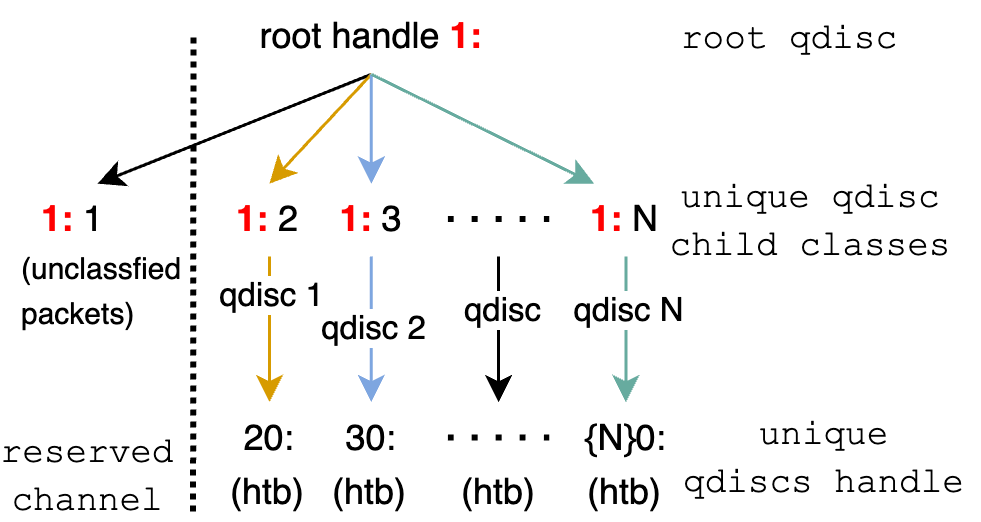}}
  \hfill
  \subfloat[Classful \texttt{qdisc} for bandwidth shaping.\label{fig:qdisc_bandwidth}]{
    \includegraphics[width=.32\linewidth]{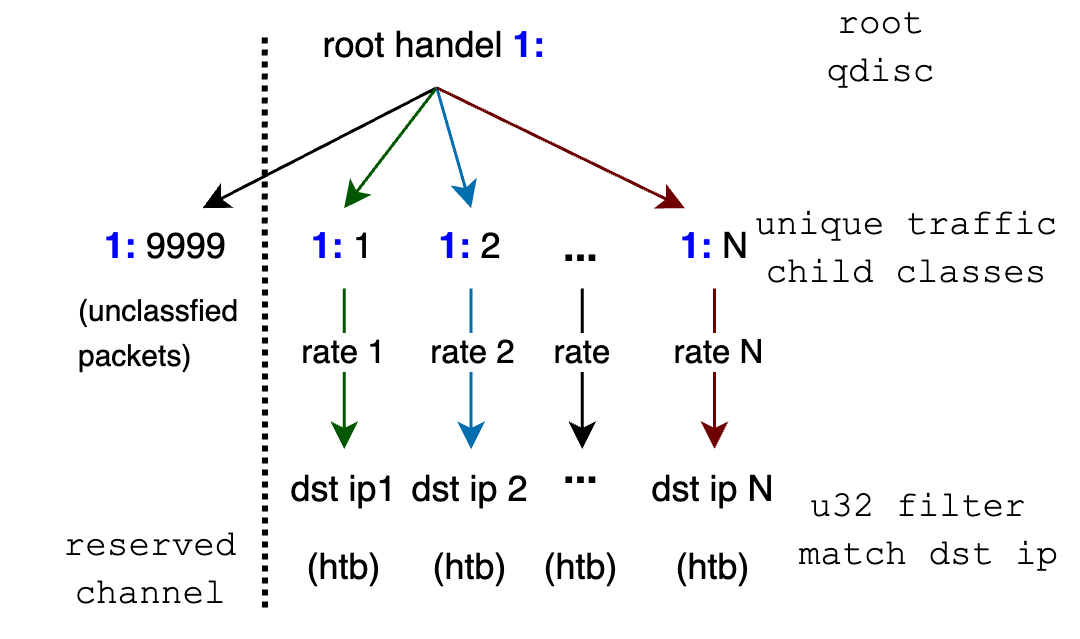}}

  \caption{The proposed method of destination-oriented networking emulation for (a) Packets are enqueued and dequeued according to classful disciplines.
           (b) Destination-specific delay emulation.
           (c) Destination-specific bandwidth shaping.}
  \label{fig:qdisc_classification}
\end{figure*}

\subsection{Linux Traffic Control Primitives}
Our emulation logic uses the Linux traffic control subsystem \cite{hubert2002linux_traffic}. At the kernel level, we rely on queueing disciplines, classes, and packet classifiers; at the user level, we program these primitives together with the \texttt{tc} utility. Specifically, the primitives that we mainly used are: (i) \textbf{qdisc (queueing discipline):} Defines how packets are enqueued and dequeued on a network interface; (ii) \textbf{class:} A logical sub-queue within a classful qdisc, and each class can have its own rate and ceiling parameters and can host a child qdisc; (iii) \textbf{netem qdisc:} A classless queueing discipline used to emulate network impairments such as delay, jitter, loss, and reordering; (iv) \textbf{u32 filter:} A general-purpose packet classifier that matches fields in the packet header (e.g., source/destination IP address, ports, protocol) by applying mask-and-shift operations on 32-bit words; (v) \textbf{\texttt{tc} utility (traffic control):} A user-space utility from the \texttt{iproute2} suite that configures the kernel traffic control subsystem. It installs, updates, and removes qdiscs, traffic classes, and filters on network interfaces, enabling emulation of bandwidth caps, delay, jitter, or loss.

In \texttt{iDynamics}, we combine a classful \texttt{htb} (Hierarchical Token Bucket) root qdisc, per-destination classes with optional \texttt{netem} children, and \texttt{u32} filters to (i) inject configurable cross-node delays and (ii) shape available bandwidth per destination, while keeping non-experimental traffic (e.g., traffic to the control plane or the public Internet) unaffected.

\subsection{Emulator: Destination-oriented Design}
\subsubsection{Emulation of Customized Delays}
Emulating diverse cross‐node delays among cluster nodes is crucial for evaluating the robustness of various scheduling policies in the cloud‐edge continuum. However, implementing customized communication delays from a single source node to multiple destination nodes poses significant challenges. To the best of our knowledge, as discussed in related work \cite{FIRM_MS, delay_ms_TMC22, delay_edge_SPE22, Adaptive_ms_IPDPS21, adaptive_ms_cloudEdge_TPDS21}, none of the existing studies has proposed an efficient method (i.e., one that is both simple and rapid) for injecting tailored communication delays in a controllable manner. In practical cloud‐edge continuum environments, these approaches generally exhibit two main limitations: (1) the use of uniform communication delays from a single source node to all destination nodes prevents differentiation between node pairs, as delays from the source to all other nodes are identical; and (2) other networking services that do not involve the correlated nodes suffer degradation because the injected delays affect all outgoing traffic.

To address these limitations, we propose a destination-oriented emulation method that classifies packets using filters to distinguish egress packets and directs them based on their IP destinations (see Figure~\ref{fig:qdisc_classification}). Additionally, we reserve an extra channel for default packet transmission without injected delays, ensuring that the performance of other services remains unaffected, for example, the packets from/to the control-plane node and outside internet services like Google. We implemented this scheme using the traffic control primitives of qdiscs (queuing disciplines) and \texttt{htb} in Linux. The packet egress structure in Figure~\ref{fig:enqueueDequeue} shows where the per-destination classes are attached before the delay and bandwidth subclasses in Figures~\ref{fig:qdisc_delay} and~\ref{fig:qdisc_bandwidth}. To emulate cross-node delays more realistically, we incorporate several factors: the base latency \textit{bl} (representing the ideal minimal latency); the maximum additional latency \textit{mal}; the distance factor \textit{df} (accounting for latency due to physical separation); and the congestion factor \textit{cf} (which simulates network uncertainties induced by traffic congestion). Here, $\Call{RandUni}{0, mal}$ denotes the additional delay generated according to a random uniform distribution between $0$ and $mal$. Thus, the emulated communication delay from node \(i\) to node \(j\) in a cluster of \(N\) cloud--edge nodes is given by the following equation:

\begin{equation}
\text{delay}^{j}_{i} = \left\lfloor \left( bl + \Call{RandUni}{0, mal} \times \frac{\lvert i - j \rvert}{N} \right) \times cf \right\rfloor
\label{eq:delay_ij}
\end{equation}

\subsubsection{Emulation of Customized Bandwidths}
Shaping customized available bandwidths between different cloud--edge node pairs is also crucial for creating dynamic networking conditions, thereby enhancing the generalizability of evaluated scheduling policies. Prior research---such as the work presented in \cite{Kollaps_EuroSys20} and \cite{THUNDERSTORM_SRDS19}---has demonstrated the feasibility of emulating dynamic network conditions, including bandwidth variations, using decentralized emulation techniques. However, these approaches typically use a global or node-level configuration that does not support fine-grained per-destination control. In contrast, our proposed approach leverages traffic control primitives, including unique traffic classes and the u32 filter, to match different IP destinations. As illustrated in Figure~\ref{fig:qdisc_bandwidth}, this approach enables the emulation of distinct bandwidth settings for different destination nodes, even when originating from the same source cluster node. This fine-grained control is essential for accurately emulating heterogeneous network conditions in cloud--edge environments.

\subsection{Measurer: Distributed Agent Design}
\label{subsec:measurer_agent}
In practical cloud--edge computing environments, fluctuating network conditions significantly impact microservice performance and SLA compliance, including varying delays and bandwidths between nodes. High latency between cluster nodes negatively affects microservice communication, while dynamically changing bandwidth can lead to congestion, increased packet loss, and potential SLA violations. To accurately capture these dynamic cross-node conditions, we introduce the \textbf{Measurer}, a distributed agent-based measurement module within the \textit{Networking Dynamics Manager} component of \texttt{iDynamics} (Figure~\ref{fig:iDynamics_framework}).

\begin{figure}[t]
    \centering
    \includegraphics[width=0.8\linewidth]{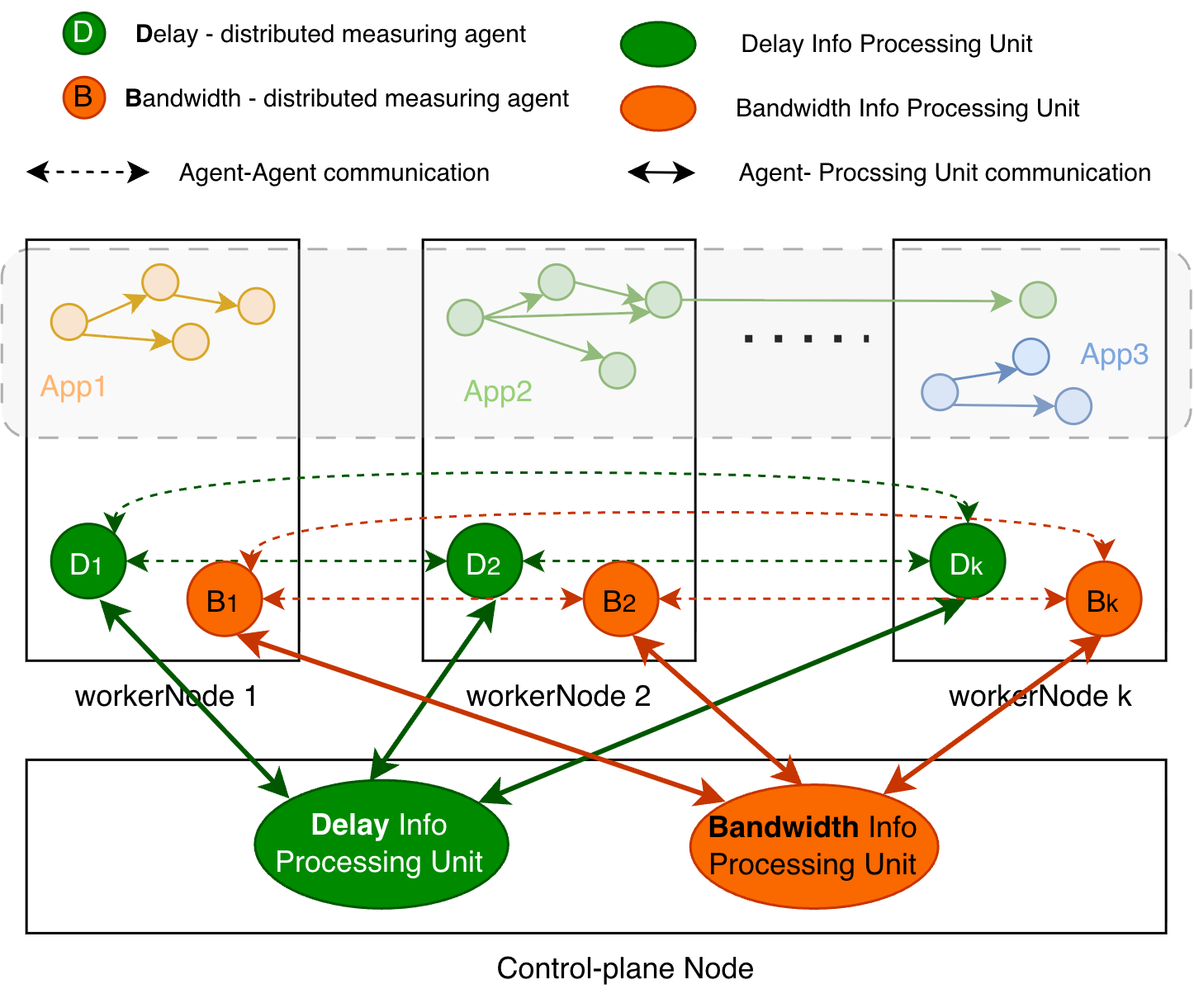}
    \caption
    {Design of \textit{Measurer} for delay and bandwidth measurement through distributed agents across cloud--edge nodes.}

    \label{fig:measure_agents}
\end{figure}

\textbf{Design Overview:} The \textbf{Measurer} uses a unified approach to capture cross-node delays and bandwidths via distributed measurement agents. Although delay and bandwidth measurements follow a similar structure, we illustrate the design using delay measurement for clarity. The module consists of a centralized information processing unit and distributed lightweight agents. The centralized processing unit maintains minimal connections with agents and efficiently aggregates the collected data. The distributed agents run as lightweight containers deployed across cluster nodes, dedicated to measuring and reporting network metrics. We implement an automatic scaling mechanism for these agents to ensure robustness and adaptability during cluster scaling. When new nodes join the cluster, corresponding agents are automatically instantiated, ensuring consistent and accurate measurements. Conversely, when nodes are removed, their associated agents are gracefully terminated.

\textbf{Implementation Details:} The centralized information processing unit is implemented as a plugin running on the control-plane node, collecting measured metrics from distributed agents via standard TCP communications. The distributed agents are deployed as Kubernetes DaemonSet pods, ensuring each node automatically hosts a dedicated measurement pod. These agents continuously measure and report cross-node delays and bandwidths, dynamically adjusting their presence in response to cluster scaling events, as depicted in Figure~\ref{fig:measure_agents}.

\textbf{Overhead Analysis:}
Regarding the overhead of measurement in a real cloud--edge cluster, we optimized the measurement using parallel processing. The measurement tasks for delays and bandwidths are designed to run concurrently, and each measurement result is aggregated into a result dictionary. Additionally, the distributed delay-measuring agents are lightweight and stable. Each node runs a delay-measuring agent built from the curlimages/curl image, consuming about 0.2 MiB of memory per node. Similarly, each node runs a bandwidth-measuring agent built from the networkstatic/iperf3 image, consuming around 0.84 MiB of memory per node. As the measurement tasks are designed to run in parallel as well, the measuring speed can be adaptively tuned by increasing or decreasing the concurrency, depending on the current load levels in the cluster.


\section{Scheduling Policy Extender}
\label{sec: Policy_Extender}
The \emph{Scheduling Policy Extender} provides the abstraction and tooling necessary to design, implement, and evaluate customized scheduling policies on top of \texttt{iDynamics}. By extending the provided interfaces and leveraging utility functions, users can efficiently develop tailored scheduling strategies that meet the diverse performance, scalability, and reliability requirements of microservice-based applications in cloud--edge environments.

\subsection{Policy Customization Interface}

\begin{figure*}[t]
    \centering
    \includegraphics[width=\linewidth]{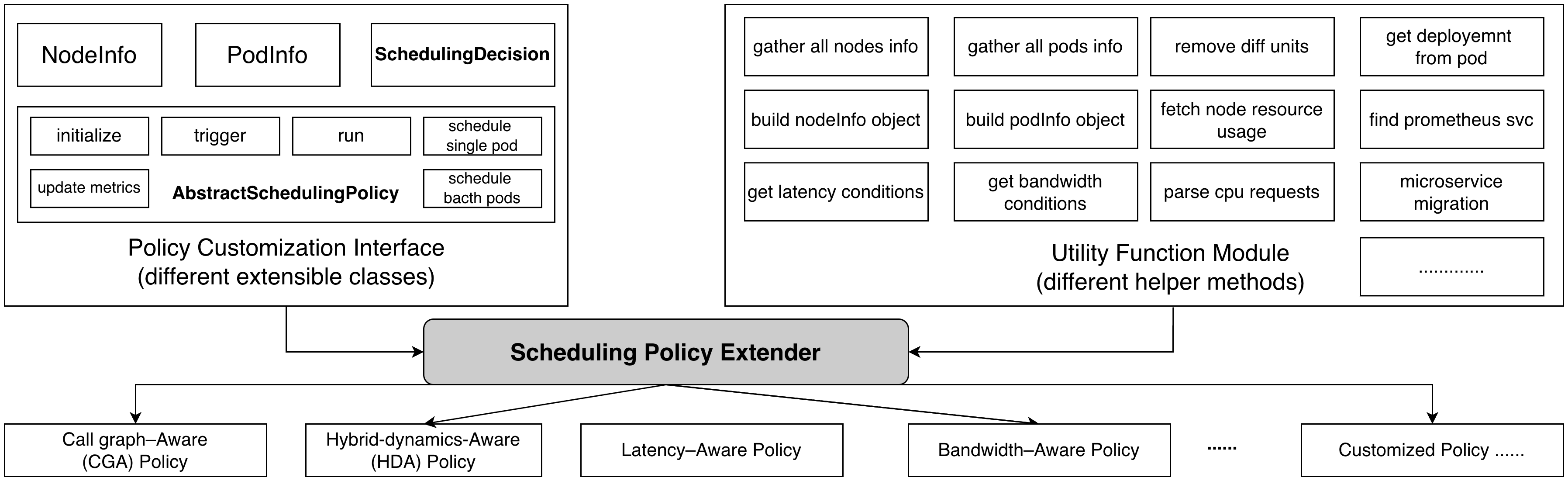}
    \caption{Overview of the \textbf{Scheduling Policy Extender}, highlighting extensible classes, utility functions, and example policies.}
    \label{fig:schedulingExtender}
\end{figure*}

As illustrated in Figure \ref{fig:schedulingExtender}, the policy interface is defined by the abstract class SchedulingPolicy, which establishes the fundamental structure for creating scheduling strategies. This abstract class includes predefined attributes and abstract methods essential for scheduling decisions. Scheduling decisions are encapsulated into combined NodeInfo and PodInfo objects. The NodeInfo class includes node-specific attributes, such as resource capacities (e.g., CPU and memory) and network characteristics (e.g., latency and bandwidth). The PodInfo class contains essential pod details, including resource requests, limits, and SLA requirements (e.g., throughput, response time), and is extensible to accommodate additional metrics. With this interface, researchers and cloud practitioners can implement their customized scheduling methods by overriding these logics, such as single-pod scheduling, batch-pod scheduling, and update-metrics methods triggered by adaptive scheduling signals.

\subsection{Utility Function Module}
The utility function module complements the policy interface, providing ready-to-use functions that simplify and accelerate the extension of custom policies. At the node level, it includes functions for collecting real-time node resource usage and network metrics, transforming them into structured objects compatible with scheduling algorithms. At the pod level, utility functions help manage pod resource specifications, SLA requirements, and metric unit conversions. Furthermore, additional helper functions utilize monitoring tools such as Prometheus to gather comprehensive cluster-level metrics and network conditions. By leveraging these pre-built utility functions, practitioners and researchers can rapidly and effectively develop sophisticated scheduling policies tailored to specific operational contexts.


\subsection{Examples of Customized Policies}
\label{subsec: examplePolicies}
To demonstrate the expressiveness of the policy interface, we design below policies that can be implemented to deal with cloud--edge dynamics:

\textbf{CGA (Call-graph--Aware) Policy:} This policy leverages the call graph and traffic stress information from the \textit{Graph Dynamics Analyzer}. Microservice pairs with high traffic stress are preferentially co-located on the same node or on nodes with low inter-node latency. This reduces cross-node communication for hot dependencies and mitigates SLA violations caused by communication bottlenecks.



\textbf{HDA (Hybrid-dynamics--Aware) Policy:} This policy jointly considers call graph structure, traffic stress, and cross-node delays. It formulates microservice placement as a Service-Node Mapping Problem: microservices with high mutual traffic are mapped to nodes with low mutual delays to minimize total communication cost. This policy illustrates how multiple dynamics can be integrated within a single optimization formulation.

\textbf{Policy cost models.}
To make the example policies more concrete, we briefly formalize their
underlying objectives using the notation introduced in
Sections~\ref{sec: Graph_Dynamics_Analyzer}
and~\ref{sec: Networking_Dynamics_Manager}. For monitoring interval
\(I_t\), the Graph Dynamics Analyzer provides the active weighted runtime
call graph \(G_t=(\mathcal{M}_t,E_t,w_t)\), where \(\mathcal{M}_t\) is
the set of running microservices, \(E_t\) is the set of observed directed
UM--DM edges, and \(w_t(\mu,\sigma)=\operatorname{stress}_t(\mu,\sigma)\)
is the traffic stress defined in Section~\ref{sec: Graph_Dynamics_Analyzer}.
Let \(\mathcal{N}\) denote the set of worker nodes. The Networking Dynamics
Manager provides the current cross-node delay \(\text{delay}^{j}_{i}\)
from node \(i\) to node \(j\), following
Equation~\ref{eq:delay_ij}. A placement for interval \(I_t\) is a mapping
\(\pi_t: \mathcal{M}_t \rightarrow \mathcal{N}\) that assigns each
microservice to a worker node.

For \emph{CGA Policy}, we can define a simple affinity score that quantifies how preferable a candidate node $n \in \mathcal{N}$ is for placing a microservice $\mu$, given a tentative placement $\hat{\pi}_t$ of its communication
partners:
\begin{equation}
\operatorname{score}^{\mathrm{CGA}}_{t}(\mu,n)
=
\sum_{\substack{(\mu,\sigma)\in E_t\\ \hat{\pi}_t(\sigma)=n}}
w_t(\mu,\sigma)
+
\sum_{\substack{(\sigma,\mu)\in E_t\\ \hat{\pi}_t(\sigma)=n}}
w_t(\sigma,\mu)
\label{eq:policy1_score}
\end{equation}
Intuitively, \(\operatorname{score}^{\mathrm{CGA}}_{t}(\mu,n)\) is high
when many of \(\mu\)'s heavily communicating upstream or downstream
neighbors in \(E_t\) are already placed on \(n\). In our implementation,
CGA Policy greedily assigns or migrates microservices to
nodes that maximize \(\operatorname{score}^{\mathrm{CGA}}_{t}(\mu,n)\),
subject to per-node resource constraints (e.g., CPU and memory capacity),
thereby co-locating hot dependencies and reducing cross-node traffic.

For \emph{HDA Policy}, we introduce an explicit
Service--Node Mapping objective that combines traffic stress and cross-node
delays. The total communication cost of a placement \(\pi_t\) during
interval \(I_t\) is defined as:
\begin{equation}
C_t(\pi_t)
= \sum_{(\mu,\sigma) \in E_t}
w_t(\mu,\sigma) \cdot
\text{delay}^{\pi_t(\sigma)}_{\pi_t(\mu)}
\label{eq:service_node_mapping}
\end{equation}
The ideal service placement then solves
\(\min_{\pi_t} \; C_t(\pi_t)\), subject to per-node resource-capacity
constraints. In other words, active UM--DM edges with high runtime traffic
stress are mapped to low-delay node pairs to minimize the aggregate
communication cost along the current active call graph.


Note that these example policies are not meant to be exhaustive but show how \texttt{iDynamics} can support different strategies that leverage different aspects of the exposed dynamics.

\section{Performance Evaluation}
\label{sec:performance_evaluation}
This section evaluates whether \texttt{iDynamics} can
construct, emulate, measure, and expose the dynamics required for
microservice scheduling-policy evaluation. Rather than claiming that the
example policies are universally optimal, our goal is to validate the
framework mechanisms and demonstrate that they provide reproducible
inputs for policy comparison. We organize the evaluation around the
following research questions:

(i) \textbf{RQ1:} What is the overhead of Algorithm~\ref{alg:call_graph_builder} while building the active runtime call graphs?
(ii) \textbf{RQ2:} How accurately and efficiently can \texttt{iDynamics}
inject and measure cross-node latency and bandwidth dynamics?
(iii) \textbf{RQ3:} Besides random matrices of networking states, can \texttt{iDynamics} replay trace-driven and burst-correlated network dynamics?
(iv) \textbf{RQ4:} Can \texttt{iDynamics} represent continuous call-graph evolution under overlapping request mixes?
(v) \textbf{RQ5:} Can \texttt{iDynamics} support different microservice-based applications and evaluation of call-graph-aware (CGA) and hybrid-dynamics-aware (HDA) policies?

\subsection{Cloud--Edge Testbed Setup}
\label{subsec:testbed_setup}

\textbf{Cluster setup:}
We implemented and evaluated \texttt{iDynamics} on a Kubernetes-based
cloud--edge testbed~\cite{kubernetes}. The cluster
contains total 46 nodes, with one control-plane node and 45 worker nodes. The control-plane node has 16 CPU cores and 64\,GiB of RAM. Among the worker nodes, 10 worker nodes have 4 CPU cores and 16\,GiB of RAM, while the remaining 35 worker nodes have 2 CPU cores and 9\,GiB of RAM. All worker and control-plane CPUs belong
to the same AMD EPYC Milan processor family (\texttt{x86\_64}). The cluster nodes communicate through a high-speed virtual network, with \texttt{iperf3} measurements
showing approximately 16--22\,Gbps of available bandwidth. The software
stack uses Kubernetes v1.36.0, Calico v3.32.0 as the CNI plugin, Istio
v1.30.0 as the service mesh, and CRI-O v1.36.0 as the container runtime.
All nodes run Ubuntu 22.04.2 LTS with Linux kernel version 5.15.0.

The nodes are virtual machines hosted on the university's
dedicated research cloud. The baseline inter-node latency is low
(typically 0.2--1\,ms in ICMP ping tests), so we do not treat the raw
testbed as a naturally geo-distributed edge environment. Instead,
\texttt{iDynamics} injects controlled cross-node delay and bandwidth
constraints to emulate cloud--edge networking conditions. Unless stated
otherwise, the injected conditions are updated during the experiments to
evaluate framework behavior and policy decisions under varying network
dynamics.

\subsection{Overhead Analysis of Algorithm~\ref{alg:call_graph_builder}}
\label{subsec:gda_algorithm_overhead}

To answer \textbf{RQ1}, we evaluate the overhead of
Algorithm~\ref{alg:call_graph_builder}, which constructs the active runtime
call graph used by the GDA (Graph Dynamics Analyzer).
Table~\ref{tab:gda_scalability_overhead_revised} reports both real benchmark
deployments and synthetic service scales in the current cluster testbed. In the
real Online Boutique~\cite{onlineBoutique_gcp_microservices_demo} and Social
Network~\cite{deathStarBench_ASPLOS19} deployments, the call graph builder
issues only two aggregate Prometheus queries instead of the logical dense
baseline $2M(M-1)$, reducing telemetry query load by 132.0$\times$ and
702.0$\times$, respectively. Their p95 total overheads are 11.405\,ms and
10.354\,ms. The local graph-build p95 is below 0.6\,ms in both cases, so the
live overhead is dominated by Prometheus aggregate-vector query latency.

The synthetic rows isolate local graph construction and
query-count scaling at larger service counts. At 1,000 services with 4,000
active edges, sparse graph construction takes 42.300\,ms p95 build time,
compared with 111.331\,ms for the measured dense pair-scan loop and 1,998,000
logical dense queries. At 50,000 services with 200,000 active edges, the sparse
builder still uses two telemetry queries and materializes the graph in
2696.727\,ms, while the dense logical baseline would require 4,999,900,000
pairwise queries. Overall, these results show that
Algorithm~\ref{alg:call_graph_builder} bounds monitoring-plane overhead by
replacing the dense $2M(M-1)$ pairwise query pattern with two aggregate telemetry
queries. The remaining local build cost scales with the materialized sparse
graph, i.e., the service/workload vertices and active directed edges, rather
than all possible service pairs.

\begin{table}[t]
\centering
\caption{Overhead of Algorithm~\ref{alg:call_graph_builder} on real benchmarks and synthetic service scales.}
\label{tab:gda_scalability_overhead_revised}
\scriptsize
\setlength{\tabcolsep}{1.1pt}
\begin{threeparttable}
\begin{tabular}{llrrrrrrr}
\toprule
\textbf{Case} & \textbf{Mode} & \textbf{Svc.} & \textbf{Edges} & \textbf{Queries} & \textbf{Q-red.} & \textbf{Query} & \textbf{Build} & \textbf{Total} \\
 & & & & & & \textbf{p95(ms)} & \textbf{p95(ms)} & \textbf{p95(ms)} \\
\midrule
\multicolumn{9}{l}{\emph{Real benchmarks}} \\
Online Boutique & sparse & 12 & 15 & 2 & 132.0$\times$ & 11.035 & 0.369 & 11.405 \\
Social Network & sparse & 27 & 24 & 2 & 702.0$\times$ & 9.897 & 0.532 & 10.354 \\
\hline
\multicolumn{9}{l}{\emph{Synthetic service scale}} \\
1,000 services & dense & 1K & 4K & 1,998K & 1.0$\times$ & -- & 111.331 & -- \\
1,000 services & sparse & 1K & 4K & 2 & 999.00K$\times$ & -- & 42.300 & -- \\
5,000 services & sparse & 5K & 20K & 2 & 25.00M$\times$ & -- & 211.071 & -- \\
10,000 services & sparse & 10K & 40K & 2 & 99.99M$\times$ & -- & 472.450 & -- \\
20,000 services & sparse & 20K & 80K & 2 & 399.98M$\times$ & -- & 974.990 & -- \\
50,000 services & sparse & 50K & 200K & 2 & 2.50B$\times$ & -- & 2696.727 & -- \\
50,000 services & dense & 50K & 200K & $\sim$ 5B & 1.0$\times$ & -- & -- & -- \\
\bottomrule
\end{tabular}
\begin{tablenotes}[flushleft]
\footnotesize
\item[] \emph{Notes:} All time columns report p95 latency in milliseconds. \textit{Svc.} denotes the deployed service/workload vertices used by the call-graph builder; in real benchmark rows, it counts Kubernetes deployments with positive replicas in the evaluated namespace. For real benchmark rows, \textit{Query} is Prometheus aggregate-vector latency, \textit{Build} is the local wall-clock time to materialize the call graph from returned telemetry edges, and
\textit{Total} is their sum. For synthetic rows, Query and Total are not reported because no live Prometheus query is issued; Build reports sparse graph materialization time for sparse rows and dense pair-scan time for the measured 1,000-service dense baseline. Q-red. is the query-count reduction relative to the logical dense pairwise baseline $2M(M-1)$. Dense synthetic timing is measured only at 1,000 services; larger dense rows are query-count baselines only.
\end{tablenotes}
\end{threeparttable}
\end{table}



\subsection{Validation of Networking Dynamics Manager}
\label{subsec:networking_validation}

\begin{table}[t]
\centering
\caption{Accuracy of cross-node network emulation.}
\label{tab:ndm_accuracy_summary}
\scriptsize
\setlength{\tabcolsep}{1pt}
\begin{threeparttable}
\begin{tabular}{@{}lrrrrr@{}}
\toprule
Metric & Pairs & Mean & Median & p95 & Max \\
\midrule
Delay abs. error (ms)          & 72 & 0.274 & 0.155 & 0.852 & 1.210 \\
Bandwidth abs. error (Mbit/s) & 72 & 21.85 & 14.00 & 89.45 & 214.00 \\
Bandwidth rel. error (\%)     & 72 & 4.28  & 3.00  & 13.66 & 27.47 \\
\bottomrule
\end{tabular}
\begin{tablenotes}[flushleft]
\item[] \emph{Notes:} Statistics are computed over all directed worker-to-worker pairs in the 9-worker validation matrix, excluding self-pairs. Delay error is the absolute difference between injected and measured delay. Bandwidth error is the absolute or relative difference between saturated and measured bandwidth. Reserved channels were not delay-injected; measured delay remained 0.21--0.71~ms to the control-plane node and 14.8--15.3~ms to external \texttt{google.com}.
\end{tablenotes}
\end{threeparttable}
\end{table}

To answer \textbf{RQ2}, we validate the design of \emph{Networking Dynamics Manager}, which is responsible for injecting and measuring cross-node latency and bandwidth.

\textbf{Latency emulation accuracy:}
We validate the destination-oriented delay-emulation scheme in Figure~\ref{fig:qdisc_delay} by injecting distinct source--destination delays between worker nodes and measuring the resulting delays using the distributed measuring agents. As shown in Table~\ref{tab:ndm_accuracy_summary}, the mean absolute delay error is only 0.274~ms, with a median error of 0.155~ms and a p95 error of 0.852~ms. Even the maximum observed error is 1.210~ms. These results show that the measured delays closely track the injected values across heterogeneous node pairs. The remaining deviations are small and are expected in a virtualized Kubernetes environment due to operating-system scheduling, packet-processing variability, and measurement noise.


Importantly, the reserved channels are not affected by the injected worker-to-worker delays. Across the validation run, measured delays from worker nodes to the control-plane node remain between 0.21 and 0.71~ms, while delays to \texttt{google.com} remain between 14.8 and 15.3~ms. This confirms that the destination-oriented design can inject delays for selected worker-to-worker paths while preserving default communication channels for non-targeted traffic.

\textbf{Bandwidth shaping accuracy:}
We similarly validate bandwidth shaping by saturating worker-to-worker paths using \texttt{iperf3}-based pods and comparing the measured throughput against the configured saturated bandwidth. Table~\ref{tab:ndm_accuracy_summary} shows that the mean absolute bandwidth error is 21.85~Mbit/s, with a median error of 14.00~Mbit/s. In relative terms, the mean error is 4.28\% and the median error is 3.00\%, indicating that the measured bandwidth generally follows the configured target closely. The p95 relative error is 13.66\%, while the maximum relative error is 27.47\%, showing that a small number of saturated paths exhibit larger deviations. This is expected for short-window bandwidth measurements in a shared virtualized testbed, where transient contention and TCP dynamics can affect saturated throughput. Overall, the results are sufficient for creating differentiated bandwidth regions, such as high-bandwidth and low-bandwidth node pairs, for controlled scheduling experiments.


Overall, Table~\ref{tab:ndm_accuracy_summary} shows that \texttt{iDynamics} can accurately emulate and measure heterogeneous cross-node network conditions. The low delay error confirms the effectiveness of destination-specific latency injection, while the bandwidth results show that the framework can reproduce differentiated bandwidth conditions with acceptable measurement variability. These capabilities provide the foundation for evaluating networking-related scheduling policies under controlled and repeatable cloud--edge dynamics.

\subsection{Trace-driven and Burst-correlated Network Dynamics}
\label{subsec:trace_dynamics_results}

To answer \textbf{RQ3}, we evaluate whether the \textit{Networking Dynamics Manager} can expose time-varying network states beyond independent distance-random matrices (used by validations in subsection~\ref{subsec:networking_validation}). The distance-random provider (Equation~\ref{eq:delay_ij}) is useful as a controllable stress baseline, but independent random samples do not preserve two important properties of practical cloud--edge networks: \textbf{temporal persistence}, where congestion lasts for multiple measurement intervals, and \textbf{spatial correlation}, where multiple paths are affected by a shared bottleneck or routing event. We therefore extend the \textit{Networking Dynamics Manager} with a \texttt{NetworkTraceProvider} abstraction that decouples network-state generation from live traffic-control application.

A provider emits a sequence of network-state frames $F_k = (L_k, B_k)$, where \(L_k \in \mathbb{R}^{N\times N}\) is a directed one-way
latency matrix and \(B_k \in \mathbb{R}^{N\times N}\) is a directed bandwidth matrix for the \(k\)-th time step. Diagonal entries are ignored because self-communication is not shaped. The framework supports three providers: (i) a \textbf{distance-random provider},
which is retained as the synthetic baseline; (ii) a \textbf{CSV-replay provider}, which replays archived time-indexed \(N\times N\) latency and bandwidth matrices; and (iii) a \textbf{burst-correlated provider}, which generates latency spikes and bandwidth drops with configurable temporal correlation, spatial correlation, burst probability, burst duration, and jitter.

\begin{table}[t]
\centering
\caption{Network trace generation and replay statistics.}
\label{tab:network_trace_provider_metrics}
\scriptsize
\setlength{\tabcolsep}{0.5pt}
\renewcommand{\arraystretch}{1.04}
\begin{threeparttable}
\begin{tabular*}{\columnwidth}
{@{\extracolsep{\fill}}lrrr@{\hspace{0.7em}}lrrr@{}}
\toprule
\multicolumn{4}{c}{\textbf{Latency}} &
\multicolumn{4}{c}{\textbf{Bandwidth}} \\
\cmidrule(r{0.7em}){1-4}
\cmidrule(l){5-8}
\textbf{Metric} &
\textbf{Dist.} &
\textbf{Burst} &
\textbf{CSV} &
\textbf{Metric} &
\textbf{Dist.} &
\textbf{Burst} &
\textbf{CSV} \\
\midrule
p50 (ms)       & 14.02  & 18.59  & 18.59  &
p50 (Mbit/s)   & 517.65 & 368.33 & 368.33 \\
p95 (ms)       & 42.83  & 45.12  & 45.12  &
p95 (Mbit/s)   & 776.48 & 574.46 & 574.46 \\
p99 (ms)       & 58.83  & 63.63  & 63.63  &
p99 (Mbit/s)   & 795.06 & 615.24 & 615.24 \\
CV             & 0.692  & 0.543  & 0.543  &
CV             & 0.333  & 0.286  & 0.286 \\
Peak/med.      & 5.38   & 3.87   & 3.87   &
Peak/med.      & 1.54   & 1.72   & 1.72 \\
Lag-1          & -0.056 & 0.946  & 0.946  &
Lag-1          & -0.015 & 0.958  & 0.958 \\
Spatial        & -0.024 & 0.067  & 0.067  &
Spatial        & -0.019 & 0.444  & 0.444 \\
Burst duration (s) & 6.37   & 27.95  & 27.95  &
Burst dur. (s) & 6.17   & 14.23  & 14.23 \\
\bottomrule
\end{tabular*}
\begin{tablenotes}[flushleft]
\item[] \emph{Notes:} \raggedright
Dist., Burst, and CSV denote \textbf{distance-random provider}, \textbf{burst-correlated provider}, and
\textbf{CSV-replay provider}, respectively. Statistics are computed over all directed non-self node pairs and all time steps. CV denotes coefficient of variation; Lag-1 denotes mean temporal autocorrelation; Spatial denotes correlation across directed node pairs. CSV-replay provider uses the archived burst-correlated matrix sequence and reproduces the same processed metrics.
\end{tablenotes}
\end{threeparttable}
\end{table}

Table~\ref{tab:network_trace_provider_metrics} shows that the
distance-random baseline has near-zero temporal autocorrelation
for both latency and bandwidth, indicating that consecutive samples
are largely independent. By contrast, the burst-correlated provider
preserves strong temporal persistence: the lag-1 autocorrelation is
0.946 for latency and 0.958 for bandwidth. It also produces longer
network events, increasing the mean latency-burst duration from
6.37\,s to 27.95\,s and the mean bandwidth-burst duration from
6.17\,s to 14.23\,s. The bandwidth trace further exhibits spatial
correlation of 0.444, showing that bandwidth drops can affect
multiple paths together. These properties make the generated
network states more suitable for evaluating schedulers under
persistent congestion and correlated bottlenecks than independent
random matrices.


To connect the generated traces to an external public latency source, we further calibrated the burst-correlated latency generator against RIPE Atlas built-in IPv4 ping measurement~1001 toward \texttt{k.root-servers.net}~\cite{RIPE_Atlas}. RIPE Atlas is a global, open, co-operatively built internet measurement network designed to provide a real-time understanding of internet connectivity and reachability. In our evaluation, the calibration uses 2,756 valid RTT (Round-Trip Time) samples from eight contributing probes over the specified window.

\begin{table}[t]
\centering
\caption{External latency calibration against real RIPE Atlas traces.}
\label{tab:ripe_latency_calibration}
\scriptsize
\setlength{\tabcolsep}{1.5pt}
\renewcommand{\arraystretch}{1.08}
\begin{threeparttable}
\begin{tabular}{@{}lrrrrrr@{}}
\toprule
\textbf{Trace Data} &
\textbf{p50} &
\textbf{p95} &
\textbf{p99} &
\textbf{CV} &
\textbf{Peak/med.} &
\textbf{Lag-1} \\
\midrule
RIPE Atlas (\colorbox{green!30}{real}) &
22.63 & 75.72 & 79.65 & 1.86 & 116.0 & 0.066 \\
Fitted generator (\colorbox{purple!30}{fitting}) &
24.50 & 64.96 & 83.99 & 0.520 & 4.551 & 0.859 \\
\bottomrule
\end{tabular}
\begin{tablenotes}[flushleft]
\footnotesize
\item[] \emph{Notes:} The RIPE Atlas row is computed directly from
2,756 valid RTT samples from eight probes over the
2026-06-02 to 2026-06-03 window. The fitted generator row is
computed from a synthetic latency matrix sequence generated by
\textbf{burst-correlated provider}. Its parameters are selected by
grid-search to match the RIPE sample's median and tail latency
scale. The fitted generator is therefore not a replay of RIPE packets
or paths. The fitting calibration is latency-only because RIPE Atlas ping
does not provide bandwidth measurements.
\end{tablenotes}
\end{threeparttable}
\end{table}

\begin{figure}[t]
    \centering
    \includegraphics[width=0.8\linewidth]{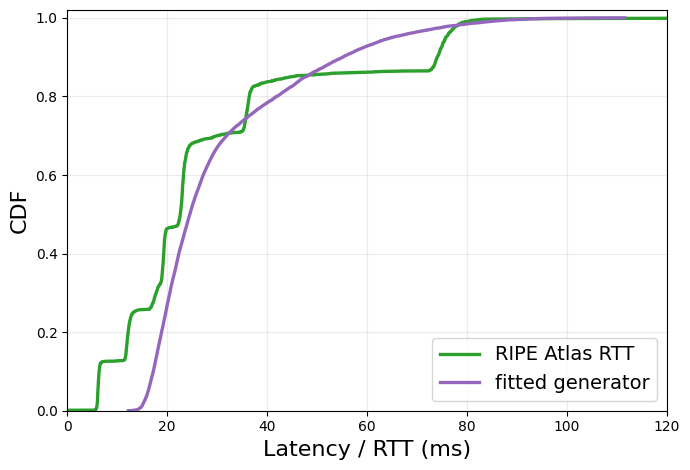}
    \caption{Empirical CDF (Cumulative Distribution Function) of RIPE Atlas RTT samples with the fitted burst-correlated latency generator.
    }
    \label{fig:label_CDF_external_calibration}
\end{figure}

Table~\ref{tab:ripe_latency_calibration} and Figure~\ref{fig:label_CDF_external_calibration} show that the fitted generator matches the main latency scale of the public RTT sample: the median is 24.504\,ms versus 22.626\,ms in the RIPE sample, and the p99 is 83.998\,ms versus 79.646\,ms. However, the generator does not reproduce all properties of the public trace. In particular, the
RIPE sample contains rare extreme outliers, reflected by a peak-to-median ratio of 116.0, whereas the fitted generator has a peak-to-median ratio of 4.551. The RIPE sample also has low lag-1 autocorrelation, while the selected generator intentionally preserves high temporal persistence. We therefore use this calibration to bound
the latency scale of the generator, not to claim that one public RTT sample represents all WAN (Wide Area Network) or cloud--edge systems. Bandwidth remains synthetic in this calibration because the public ping measurement does not provide throughput data.

Overall, the results answer \textbf{RQ3}: \texttt{iDynamics} can support various networking dynamics by replaying archived time-indexed latency/bandwidth matrices, generating burst-correlated traces with temporal and spatial structure, and calibrating latency scale against the external public RTT sample trace.


\subsection{Continuous Call-graph Evolution and Policy Evaluation}
\label{subsec:continuous_callgraph_policy_eval}

To answer \textbf{RQ4} and the policy-evaluation part of
\textbf{RQ5}, we evaluate continuous call-graph evolution with mixed request types, proportions and modes. The workload mixer keeps multiple
request classes active concurrently and changes their mix proportions under
\textbf{step}, \textbf{linear}, \textbf{sinusoidal}, and \textbf{Markov}
modes, summarized in Table~\ref{tab:request_mix_modes}. Consequently, each
monitoring interval produces an active weighted call graph whose topology and
edge weights reflect the combined effect of overlapping request classes. This
setting exercises the core framework requirement that scheduling policies
consume a stream of evolving graph and traffic loads, instead of a single
static dependency graph.


%

\begin{figure*}[t]
    \centering
    \includegraphics[width=\textwidth]{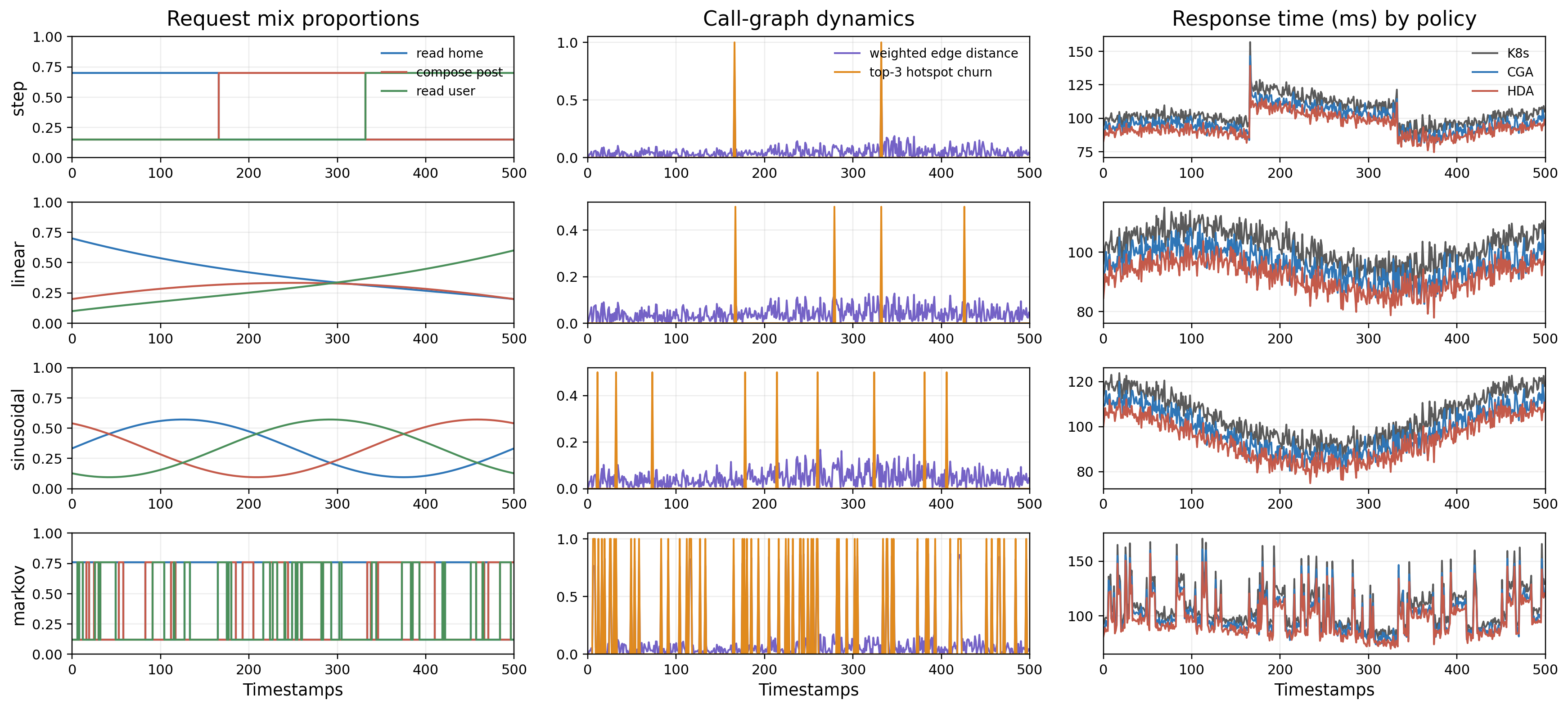}
   \caption{Continuous call-graph evolution under varying request types, proportions, and mix modes for Social Network benchmark. Columns display request-mix proportions, graph-dynamics metrics, and response times. Weighted edge distance quantifies normalized traffic-weight redistribution between adjacent call graph snapshots, while Top-3 hotspot churn reports whether the three dominant directed edges change in the graph. The evaluations are conducted at 90 QPS with 5-second interval timestamps across a 45-node scale in the testbed.}
    \label{fig:continuous_longmix_500_multimode}
\end{figure*}

\begin{table}[t]
\centering
\caption{Request-mix generation modes.}
\label{tab:request_mix_modes}
\footnotesize
\setlength{\tabcolsep}{4pt}
\renewcommand{\arraystretch}{1.15}
\begin{tabularx}{\columnwidth}{@{}l >{\raggedright\arraybackslash}X@{}}
\toprule
\textbf{Mode} & \textbf{Semantics} \\
\midrule
\mode{step} &
One request class is dominant for each segment while all other classes remain
active at baseline weight. \\
\addlinespace[1pt]
\mode{linear} &
Dominance shifts smoothly from the first request class to the last, with a
midpoint bump for interior request classes. \\
\addlinespace[1pt]
\mode{sinusoidal} &
All request classes remain active while phase-shifted sinusoidal weights change the hot request over time. \\
\addlinespace[1pt]
\mode{markov} &
A Markov schedule changes the dominant request class stochastically while
preserving nonzero baseline traffic for the rest. \\
\bottomrule
\end{tabularx}
\end{table}

We use two complementary call-graph metrics to
separate the edge-weight movement (\textit{traffic evolution}) from the changes of
dominant active edges (\textit{topology evolution}) in the call graph. Let \(E_t\) be the active edge set of the directed weighted call graph at
interval \(t\), and let \(w_t(e)\) be the nonnegative traffic-stress weight of
edge \(e\). The \textbf{weighted edge distance} is a normalized sum of absolute weight differences across all edges in the union of \(E_{t-1}\) and \(E_t\):
\begin{equation}
D_w(E_{t-1},E_t)=
\frac{\sum_{e\in E_{t-1}\cup E_t}|w_{t-1}(e)-w_t(e)|}
{\sum_{e\in E_{t-1}\cup E_t}\max(w_{t-1}(e),w_t(e))}.
\label{eq:weighted_edge_distance}
\end{equation}
Missing edges are treated as weight zero. Thus, \(D_w=0\) means adjacent call graph
snapshots have identical weighted traffic, whereas larger values mean
that traffic has shifted, appeared, or disappeared across directed service
pairs. The normalization makes the score comparable across different graph
sizes and traffic levels.

\textbf{Top-3 hotspot churn} focuses on a different perspective: whether the main
bottleneck edge candidates change in the graph as time goes by. Let \(H_t^{(3)}\) be the set of the
three highest-weight directed edges in \(E_t\). We compute
\begin{equation}
C_3(E_{t-1},E_t)=
1-\frac{|H_{t-1}^{(3)}\cap H_t^{(3)}|}
{|H_{t-1}^{(3)}\cup H_t^{(3)}|}.
\label{eq:top3_hotspot_churn}
\end{equation}
This is a Jaccard-distance score over the hotspot sets. The Jaccard distance is a metric used to quantify how dissimilar two sets are. It measures the difference between sample sets by dividing the size of the set difference by the size of the set union. Scores range from 0 (identical) to 1 (completely disjoint). In Equation~\ref{eq:top3_hotspot_churn}, a value of zero means the same three directed edges remain dominant; a high value means the heaviest traffic paths have moved to different service pairs.

\textbf{Why use these two metrics?} An \textit{active runtime call graph} can have low hotspot churn but nonzero weighted distance: the same top edges remain, but their traffic weights shift.
The call graph can also have high hotspot churn while total traffic volume is similar: the bottleneck edges move to different service pairs. That distinction is useful for scheduling-policy evaluation. A scheduler may care about total traffic redistribution, but it may care even more when the dominant edges change in the call graph, because those are the graph edges most likely to affect microservice placement, colocation, delay sensitivity, or bandwidth pressure. Therefore, these two metrics are complementary in an \textit{active runtime call graph}: weighted edge distance measures how much
traffic load changes, while Top-3 hotspot churn measures whether the most
policy-relevant hot edges change.

\begin{table}[t]
\centering
\caption{Performance and uncertainty of continuous request mixes under robustness evaluation.}
\label{tab:continuous_longmix_robustness}
\scriptsize
\setlength{\tabcolsep}{1.3pt}
\renewcommand{\arraystretch}{1.06}

\begin{threeparttable}
\begin{tabular*}{\columnwidth}
{@{\extracolsep{\fill}}rlrrrrrr@{}}
\toprule
\multirow{2}{*}{\textbf{Steps}} &
\multirow{2}{*}{\textbf{Mode}} &
\multirow{2}{*}{\textbf{Entropy}} &
\multirow{2}{*}{\shortstack{\textbf{Top-3}\\\textbf{churn}}} &
\multirow{2}{*}{\shortstack{\textbf{Active}\\\textbf{edges}}} &
\multicolumn{3}{c}{\textbf{Response time (ms)}} \\
\cmidrule(l){6-8}
& & & & &
\textbf{K8s} &
\textbf{CGA} &
\textbf{HDA} \\
\midrule

\multirow{4}{*}{200}
& Step       & 1.181 & 0.010 & 13.3 & 104.52 & 98.25  & 93.83 \\
& Linear     & 1.488 & 0.010 & 15.0 & 101.93 & 95.82  & 91.51 \\
& Sinusoidal & 1.387 & 0.023 & 13.3 & 104.78 & 98.50  & 94.07 \\
& Markov     & 1.035 & 0.141 & 11.0 & 104.60 & 98.32  & 93.91 \\
\hline
\addlinespace[1.5pt]

\multirow{4}{*}{500}
& Step       & 1.181 & 0.004 & 13.3 & 104.38 & 98.14  & 94.01 \\
& Linear     & 1.489 & 0.004 & 15.0 & 101.94 & 95.86  & 91.82 \\
& Sinusoidal & 1.387 & 0.009 & 13.3 & 104.73 & 98.47  & 94.32 \\
& Markov     & 1.035 & 0.136 & 11.2 & 107.38 & 100.97 & 96.72 \\
\hline
\addlinespace[1.5pt]

\multirow{4}{*}{1000}
& Step       & 1.181 & 0.002 & 13.3 & 104.40 & 98.16  & 93.94 \\
& Linear     & 1.489 & 0.002 & 15.0 & 102.02 & 95.93  & 91.80 \\
& Sinusoidal & 1.387 & 0.005 & 13.3 & 104.80 & 98.54  & 94.30 \\
& Markov     & 1.035 & 0.155 & 11.5 & 109.75 & 103.20 & 98.76 \\

\bottomrule
\end{tabular*}

\begin{tablenotes}[flushleft]
\item[] \emph{Notes:} \raggedright
Each metric reports the \textbf{mean value} across all evaluation
timestamps in the corresponding horizon. Entropy is the Shannon entropy of the request-mix
proportion vector,
\(H(P)=-\sum_i p_i\log_2 p_i\), with zero-probability terms omitted.
Top-3 churn and active edges characterize temporal variation in the
resulting call graph. K8s, CGA, and HDA report the measured performance
under the corresponding scheduling policies; lower values are better.
\end{tablenotes}

\end{threeparttable}
\end{table}

As shown in Figure~\ref{fig:continuous_longmix_500_multimode}, the
qualitative differences among the four evolution modes are illustrated. Step mode produces
piecewise-stable plateaus; linear and sinusoidal modes produce gradual shifts
in the request-mix vector; and Markov mode produces bursty, correlated changes
in the dominant request class. In Table~\ref{tab:continuous_longmix_robustness}, the quantitative metrics show that linear mixes have the highest mean request-mix entropy and the largest active graph, with approximately 15 active directed edges. Markov mixes have lower entropy but much higher mean Top-3 hotspot churn, reaching 0.155 at the 1000-timestamp evaluation. This indicates a narrower but less stable hot-edge set, which is precisely the type of evolving bottleneck pattern that can stress microservice placement policies. Across all modes and evaluation horizons for the three evaluated policies, HDA has the lowest same-snapshot modeled response time, followed by CGA and then K8s default. The important framework result is not that HDA is universally superior, but that the same evolving call graph can be replayed through multiple executable policy objectives over long horizons while preserving a clear evidence boundary.

\subsection{Application Generality and Compatibility Case Studies}
\label{subsec:application_generality}

To answer the application-generality part of \textbf{RQ5}, we evaluate
whether \texttt{iDynamics} can preserve a common measurement and policy
interface across applications with different request classes and application call graphs. We use two complementary benchmarks. Online
Boutique~\cite{onlineBoutique_gcp_microservices_demo} is an externally
maintained, Kubernetes-native retail application whose gRPC service graph
exercises conventional cloud microservice paths such as browsing, cart, and
checkout. It tests whether the framework can attach to a realistic benchmark
that was not designed for \texttt{iDynamics}. The second benchmark is our
CPU-only MoE (Mixture of Experts) serving microbenchmark, which deploys a
role-selected HTTP service graph with a frontend, tokenizer, router/gate,
cache, aggregator, and eight expert services. Its implemented request path is
frontend \(\rightarrow\) optional cache lookup \(\rightarrow\) tokenizer
\(\rightarrow\) request-level top-\(k\) router \(\rightarrow\) selected experts
\(\rightarrow\) aggregator \(\rightarrow\) cache update. We use it because
MoE-style inference workloads introduce dynamic expert popularity,
routing-dependent fan-out/fan-in traffic, cache effects, payload-size
variation, and batch-size variation, which are qualitatively different from
e-commerce (Online Boutique) request flows. Together, these benchmarks test adapter generality across an
external application and a controllable ML-inference-style microservice graph;
they are not intended to claim universal benchmark coverage, neural-network
MoE inference fidelity, or full GPU-aware production LLM (large language
model) serving.

\begin{figure*}[t]
\centering
\subfloat[Online Boutique under a sinusoidal request-mix schedule with
45 worker nodes and 5 replicas per microservice.]{
    \includegraphics[width=0.48\textwidth]{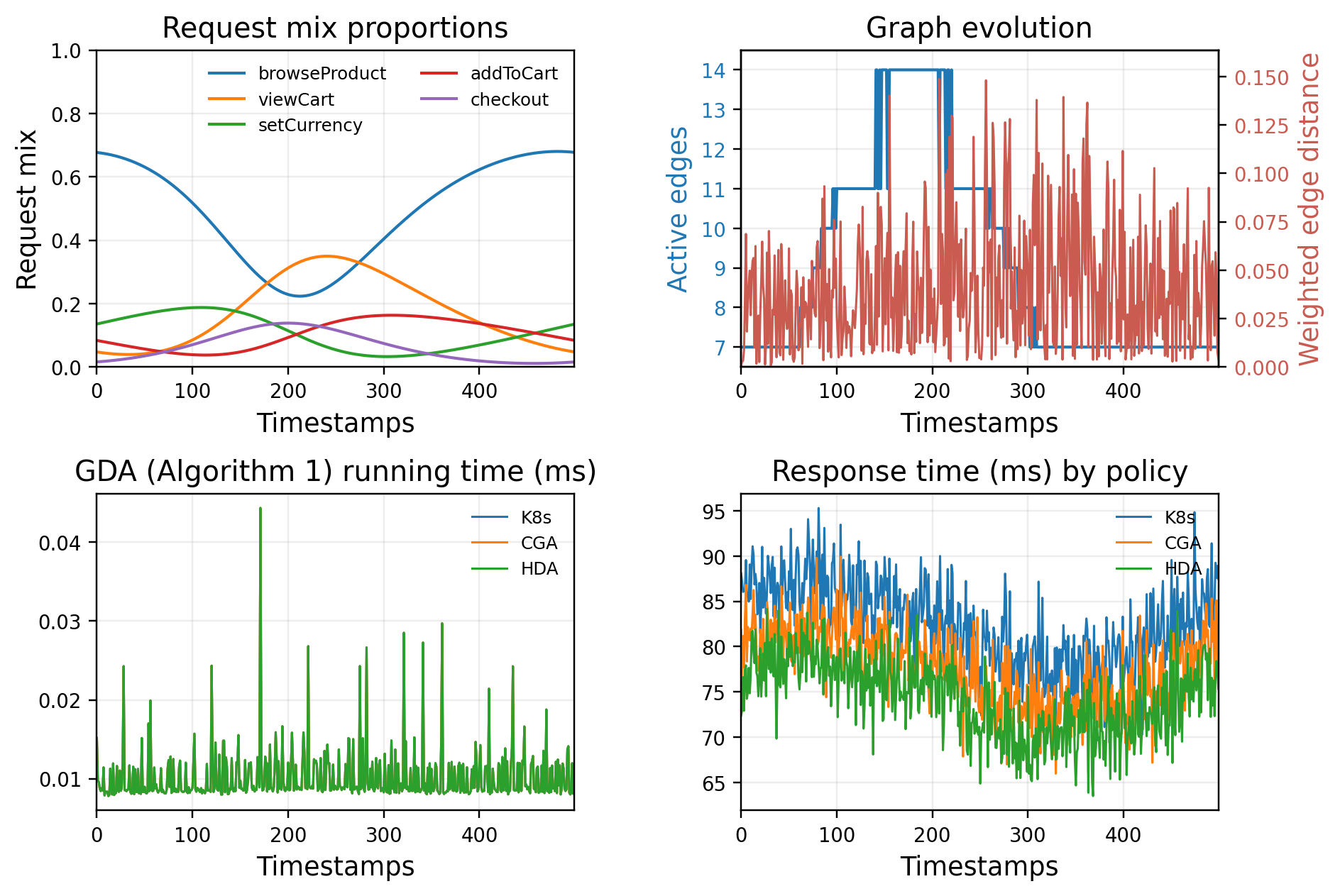}
    \label{fig:application_generality_online_boutique_timeseries}}
\hfill
\subfloat[CPU-only MoE-style microbenchmark under a Markov request-mix schedule
with 45 worker nodes and 5 replicas per microservice.]{
    \includegraphics[width=0.48\textwidth]{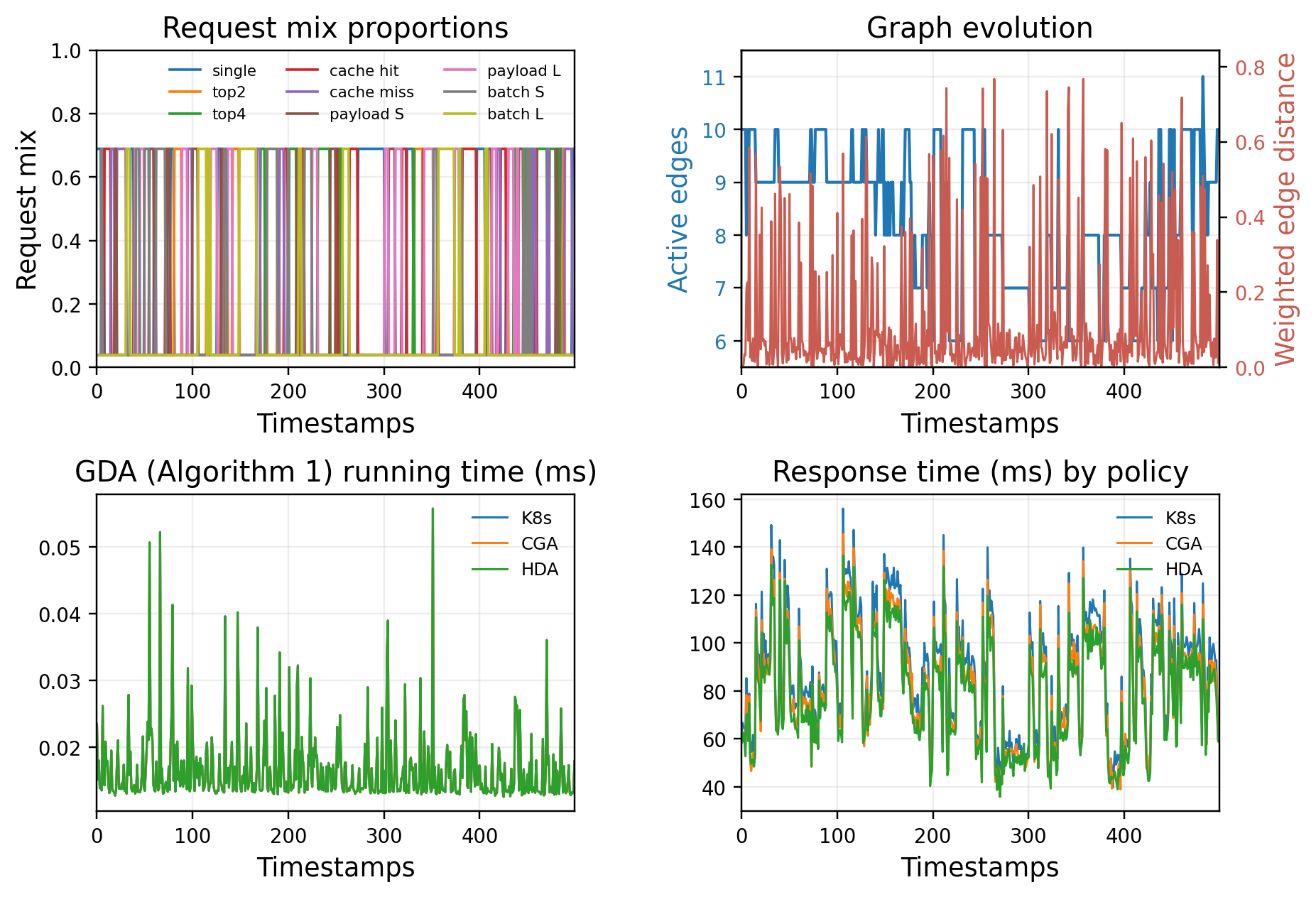}
    \label{fig:application_generality_moe_timeseries}}
\caption{Application-generality examples for Online Boutique and the
CPU-only MoE-style microbenchmark. Both panels use the same request-mix,
weighted call-graph, replay-policy, and runtime-GDA schema.}
\label{fig:application_generality_longmix}
\end{figure*}

\begin{figure}[t]
\centering
\resizebox{\columnwidth}{!}{%
\begin{tikzpicture}[
svc/.style={draw, rounded corners=1pt, align=center, minimum width=1.45cm,
minimum height=0.48cm, inner sep=2pt},
edge/.style={->, line width=0.35pt}
]
\node[svc] (load) at (0,0) {Workload\\mix};
\node[svc] (front) at (1.9,0) {Frontend};
\node[svc] (tok) at (3.8,0) {Tokenizer};
\node[svc] (router) at (5.7,0) {Router/\\gate};
\node[svc] (expert0) at (7.7,0.75) {Expert 0};
\node[svc] (expertd) at (7.7,0) {\(\cdots\)};
\node[svc] (expert7) at (7.7,-0.75) {Expert 7};
\node[svc] (agg) at (9.7,0) {Aggregator};
\node[svc] (cache) at (1.9,-1.25) {Cache/state};
\draw[edge] (load) -- (front);
\draw[edge] (front) -- (tok);
\draw[edge] (tok) -- (router);
\draw[edge] (router) -- (expert0);
\draw[edge] (router) -- (expertd);
\draw[edge] (router) -- (expert7);
\draw[edge] (expert0) -- (agg);
\draw[edge] (expertd) -- (agg);
\draw[edge] (expert7) -- (agg);
\draw[edge] (front) -- (cache);
\draw[edge] (agg) |- (cache);
\draw[edge] (agg) -- ++(1.25,0) node[right] {Response};
\node[align=center] at (5.7,-1.5) {request-level top-\(k\) expert routing};
\end{tikzpicture}}
\caption{CPU-only MoE-style serving microbenchmark used in the
application-generality study. The benchmark exposes dynamic expert popularity,
fan-out/fan-in service traffic, cache behavior, and payload effects for
placement evaluation; it is not a faithful neural-network MoE inference
engine.}
\label{fig:moe_microbenchmark_flow}
\end{figure}

Figure~\ref{fig:moe_microbenchmark_flow} summarizes the evaluated MoE
service graph, and Figure~\ref{fig:application_generality_longmix} shows that
the same request-mix, call-graph, GDA-runtime, and response time from different policies can be
produced for both benchmarks by \texttt{iDynamics}. In the Online Boutique panel
(Figure~\ref{fig:application_generality_online_boutique_timeseries}), the
sinusoidal mix changes request dominance smoothly, and the reconstructed
runtime graph changes accordingly: the active-edge count moves between narrow
and wider graph states while weighted-edge distance captures the traffic
redistribution between call graph snapshots. The response time curves from different policies exhibit a
consistent performance superiority, with HDA below CGA and CGA below K8s default over
most timestamps, while the Algorithm~\ref{alg:call_graph_builder} construction cost remains small relative to the
policy-level response-time scale. In the MoE panel
(Figure~\ref{fig:application_generality_moe_timeseries}), the Markov request mix
causes abrupt shifts among single-expert, multi-expert, cache, payload, and
batch classes; this produces sharper graph-distance bursts and larger
response-time variation because the hot expert and support-service paths move
more abruptly. The evaluations suggest that \texttt{iDynamics} does not depend on a few benchmark-specific graphs (e.g., Social Network and Online Boutique): the same dynamics adapter scheme can expose gradual
e-commerce workflow evolution and bursty expert-routing evolution through
comparable metrics and executable scheduling policies, while keeping the MoE microbenchmark claim bounded to CPU-only service-graph and placement dynamics.

\subsection{Threats to Validity and Limitations}
\label{subsec:threats_validity}

The evaluation is conducted on a controlled Kubernetes research testbed
hosted on virtual machines. The 45-worker results therefore demonstrate scalability and policy behavior in our current cluster, but they should not be interpreted as universal production performance for geographically distributed bare-metal edge deployments. Network dynamics are emulated with Linux \texttt{tc}/qdisc on a Calico overlay. Although each run records qdisc state, resets the configuration, and verifies connectivity, destination-worker shaping can affect other selected-worker overlay packets. We therefore treat the fault-injection results as controlled policy experiments, not as a proof of perfect traffic isolation.

The measurement stack also has scope limits. Service-mesh telemetry provides the UM--DM traffic labels required by GDA, but it introduces sidecar overhead that depends on mesh configuration, payload size, request path, and security settings. We quantify this overhead in our environment, rather than
claiming a mesh-independent cost. Similarly, runtime-GDA values in our evaluations measure the graph-construction path for archived snapshots; they represent live Prometheus overhead only when the corresponding experiment contains live query samples. The graph metrics used in the paper are compact summaries: weighted edge distance captures traffic redistribution, and Top-3 hotspot churn captures dominant-hotspot changes, but neither metric by itself explains the causal source of a performance change.

Finally, the benchmark and policy claims are intentionally bounded. Online Boutique supports external-application compatibility and replay-schema evidence, while the MoE case study is a CPU-only microservice benchmark for expert-skew, fan-out/fan-in, cache, payload, and placement dynamics. It does not model GPU topology, tensor parallelism, KV-cache locality, accelerator memory pressure, token-level MoE routing, or production LLM-serving schedulers. CGA and HDA are executable reference policies used to validate the Scheduling Policy Extender; they are not claimed to be state-of-the-art or universally optimal schedulers. Stronger objectives involving CPU, memory, energy, fairness, isolation, or accelerator resources can be implemented through the same interface, but require separate empirical validation.

\section{Conclusions and Future Work}
\label{sec:conclusion}

In this paper, we proposed \texttt{iDynamics}, a configurable emulation framework for evaluating microservice scheduling policies under controllable cloud--edge dynamics. \texttt{iDynamics} combines service-mesh-based active call-graph construction, destination-specific network emulation, workload
and trace adaptation, and an executable scheduling-policy interface. This design allows researchers and practitioners to replay the same evolving application and networking states across different placement objectives while preserving a clear boundary between measured evidence, replayed policy outputs, and modeled response-time estimates.

The evaluation demonstrates that \texttt{iDynamics} can construct sparse runtime call graphs efficiently, validate live \texttt{tc}-based network impairments, replay continuous request-mix and call-graph evolution over long horizons, and compare representative scheduling policies on worker pools up to
45 nodes. Additional Online Boutique and CPU-only MoE-style case studies show that the same adapter scheme can be applied beyond the Social Network benchmark, while keeping claims bounded to the evidence provided by each evaluation. Overall, the results show that \texttt{iDynamics} provides a repeatable and extensible environment for studying how scheduling policies respond to workload, application-graph, and network dynamics.

Future work will extend \texttt{iDynamics} in three directions. First, we will incorporate richer policy objectives and metrics, including CPU, memory, energy, fairness, isolation, and multi-tenant contention. Second, we will broaden benchmark coverage with additional stable event-streaming and ML-serving workloads, including GPU-aware telemetry when suitable hardware is available. Third, we will investigate lower-overhead data-plane mechanisms, such as eBPF/XDP or traffic-control offload, to support richer network behaviors at larger scales.

\textbf{Software}: \texttt{iDynamics} framework prototype as open source is available at \href{https://github.com/Cloudslab/iDynamics}{https://github.com/Cloudslab/iDynamics}

\bibliographystyle{IEEEtran}

\bibliography{reference}

\end{document}